\documentclass[pra,reprint,a4paper,showpacs,citeautoscript,floatfix]{revtex4-1}

\pdfoutput=1
\usepackage[charter]{mathdesign}
\usepackage{amsmath}
\usepackage{mathrsfs}

\usepackage[pdftex]{graphicx}
\usepackage{microtype}
\usepackage{bm}\let\vec\bm

\usepackage[svgnames]{xcolor}
\usepackage[
	colorlinks=True,linkcolor=DarkRed,citecolor=ForestGreen,urlcolor=MediumBlue,
	pdfstartview=FitH,bookmarks=False,pdfpagemode=UseNone
]{hyperref}

\def\FeTeSe{FeTe$_{0.55}$Se$_{0.45}$}

\begin{document}

\title{\boldmath Signatures of nodeless multiband superconductivity and particle-hole crossover\\in the vortex cores of \FeTeSe{}}

\author{Christophe Berthod}
\affiliation{Department of Quantum Matter Physics, University of Geneva, 24 quai Ernest-Ansermet, 1211 Geneva, Switzerland}

\date{October 24, 2018}

\begin{abstract}

Scanning tunneling experiments on single crystals of superconducting \FeTeSe{} have recently provided evidence for discrete energy levels inside vortices. Although predicted long ago, such levels are seldom resolved due to extrinsic (temperature, instrumentation) and intrinsic (quasiparticle scattering) limitations. We study a microscopic multiband model with parameters appropriate for \FeTeSe{}. We confirm the existence of well-separated bound states and show that the chemical disorder due to random occupation of the chalcogen site does not affect significantly the vortex-core electronic structure. We further analyze the vortex bound states by projecting the local density of states on angular-momentum eigenstates. A rather complex pattern of bound states emerges from the multiband and mixed electron-hole nature of the normal-state carriers. The character of the vortex states changes from hole-like with negative angular momentum at low energy to electron-like with positive angular momentum at higher energy within the superconducting gap. We show that disorder in the arrangement of vortices most likely explains the differences found experimentally when comparing different vortices.

\end{abstract}

\maketitle

\section{Introduction}

As a local and space-resolved probe of the electronic density of states, the scanning tunneling microscope (STM) \cite{Binnig-1987} has enabled detailed testing of the Bardeen--Cooper--Schrieffer (BCS) theory of superconductivity \cite{Bardeen-1957}---generalized to encompass inhomogeneous order parameters by Bogoliubov and de Gennes \cite{deGennes-1999}---especially in the interior of Abrikosov vortices \cite{Hess-1990, Gygi-1991, Kaneko-2012, Berthod-2017}. The predicted spectrum of one-electron excitations inside vortices is qualitatively different in superconductors with and without nodes in the order parameter, for in the first case the low-energy states attached to the cores are resonant with extended states outside the core and the resulting energy spectrum is continuous, while in the second case, where the superconducting state is fully gapped, the subgap vortex-core states are truly bound and the energy spectrum is discrete (assuming the vortex is isolated). Hence an experimental detection of discrete levels in vortices is sufficient to rule out nodal order parameters. Such an observation is possible only in the quantum regime, when the energy separation between the levels is larger than the energy broadening related to temperature and quasiparticle interactions. The typical energy separation is $\Delta^2/E_{\mathrm{F}}$, where $\Delta$ is the order-parameter scale and $E_{\mathrm{F}}$ is the Fermi energy \cite{Caroli-1964}. In the cuprates, the quantum regime is realized but the order parameter has nodes and the vortex-core spectrum is therefore continuous \cite{Berthod-2017}. On the other hand, most nodeless superconductors are not in the quantum regime, such that the vortex states are dense and measurements by STM show broad features inside vortices \cite{Suderow-2014}. Very recently, discrete vortex levels were discovered in \FeTeSe{} \cite{Chen-2018}.

In the quantum regime, the bound states appear as resolution-limited peaks in the local tunneling conductance. Exactly at the core center, the tunneling spectrum predicted by the Bogoliubov--de Gennes theory breaks particle-hole symmetry \cite{Caroli-1964, Bardeen-1969, Hayashi-1998, Kaneko-2012}. For an electron band (positive mass), the lowest state has energy $+\frac{1}{2}\Delta^2/E_{\mathrm{F}}$ and the electronic component of its wave function is maximal at the core center, while the hole component vanishes at the core center, such that there is no peak at energy $-\frac{1}{2}\Delta^2/E_{\mathrm{F}}$. The situation is reverted for a hole band (negative mass) \cite{Araujo-2009}, where the peak at the core center has energy $-\frac{1}{2}\Delta^2/E_{\mathrm{F}}$ and there is nothing at $+\frac{1}{2}\Delta^2/E_{\mathrm{F}}$. The second and higher-energy states have no weight at the core center. These expectations contrast with the observations made in \FeTeSe{}, where indeed some of the vortices show a single peak at positive energy at the core center, as may be expected for an electronic band, while another vortex shows an almost particle-hole symmetric pair of peaks and yet other vortices show an asymmetric spectrum with a tall peak at negative energy and a smaller peak at positive energy \cite{Chen-2018}. Moreover, some vortices show several additional peaks at the core center, where the theory predicts only one. A characteristic of \FeTeSe{}, shared with most iron chalcogenides and pnictides \cite{Liu-2015}, is a disconnected Fermi surface presenting hole pockets around the $\Gamma$ point of the Brillouin zone and electron pockets around the M points. The minimal model for \FeTeSe{} has one electron pocket and one hole pocket of nearly equal volumes. Because of this, a tempting interpretation of the experimental data would be that vortices show a mixture of electron-like and hole-like bound states. Motivated by measurements on pnictides, a number of authors have studied vortices in models featuring electron and hole bands \cite{Araujo-2009, Jiang-2009, Hu-2009, Gao-2011, Zhou-2011, Gao-2012, Wang-2015, Uranga-2016}. A comparison with \FeTeSe{} is difficult, as these works either choose parameters not in the quantum regime, or use exact diagonalization on small systems and fail to achieve the high energy resolution required in order to address discrete vortex levels. Thus, the theoretical vortex spectrum for a nearly compensated two-band metal in the quantum regime has not been computed so far.

Provided that the two-band scenario gives the correct interpretation, another mechanism must still be identified in order to explain the significant variations observed from one vortex to another. As \FeTeSe{} is microscopically disordered \cite{He-2011}, one candidate explanation rests on the variations in the local potential landscape due to random occupation of the Te/Se sites. Another possibility would be density inhomogeneities, which would locally change the balance between the electron and hole pockets. One could also adduce the irregular distribution of vortices observed in the experiments, as simulations have shown that disorder in the vortex positions impacts the local density of states (LDOS) in the cores \cite{Berthod-2016, Berthod-2017}.

A measure of the proximity to the quantum limit is provided by the product $k_{\mathrm{F}}\xi\sim E_{\mathrm{F}}/\Delta$ of the Fermi wave vector and the coherence length. The typical size of the Fermi pockets in \FeTeSe{} is \cite{Miao-2012} $k_{\mathrm{F}}\approx0.1$~\AA$^{-1}$ and the coherence length is \cite{Shruti-2015} $\xi\approx20$~\AA{}. The value $k_{\mathrm{F}}\xi\approx 2$ locates \FeTeSe{} deeper in the quantum regime compared with the only other compound where discrete vortex levels have been experimentally observed \cite{Kaneko-2012}, i.e., YNi$_2$B$_2$C. The theoretical calculations in Ref.~\onlinecite{Kaneko-2012} ($k_{\mathrm{F}}\xi=10$) as well as the earlier LDOS calculations of Ref.~\onlinecite{Hayashi-1998} ($k_{\mathrm{F}}\xi=8$) must be extended to lower values of $k_{\mathrm{F}}\xi$, before a comparison with \FeTeSe{} can be attempted. These calculations have considered an isolated vortex in a clean superconductor with free-electron-like dispersion. Further generalizations are needed in order to incorporate the electron-hole band structure and the chemical disorder present in \FeTeSe{} and for studying the effects of intervortex interactions within a disordered vortex configuration like the one observed in Ref.~\onlinecite{Chen-2018}. Our aim is to extend and complement the pioneering calculations, coming closer to the actual experimental situation of \FeTeSe{}.

We present our two-dimensional disordered tight-binding model and review its basic properties as well as our calculation methods in Sec.~\ref{sec:model}. In Sec.~\ref{sec:vortex}, we study an isolated vortex and characterize the discrete bound states in terms of their approximate angular momentum and electron-like or hole-like character. Section~\ref{sec:disorder} illustrates the changes that chemical disorder, possible density inhomogeneities, and disorder in the vortex positions can bring to vortex-core spectra. We summarize our results and present discussions and perspectives in Sec.~\ref{sec:discussion}, concluding in Sec.~\ref{sec:conclusion}. Appendices \ref{app:resolution} to \ref{app:vortices} collect additional material.

\section{Model and method}
\label{sec:model}

\begin{figure}[tb]
\includegraphics[width=0.9\columnwidth]{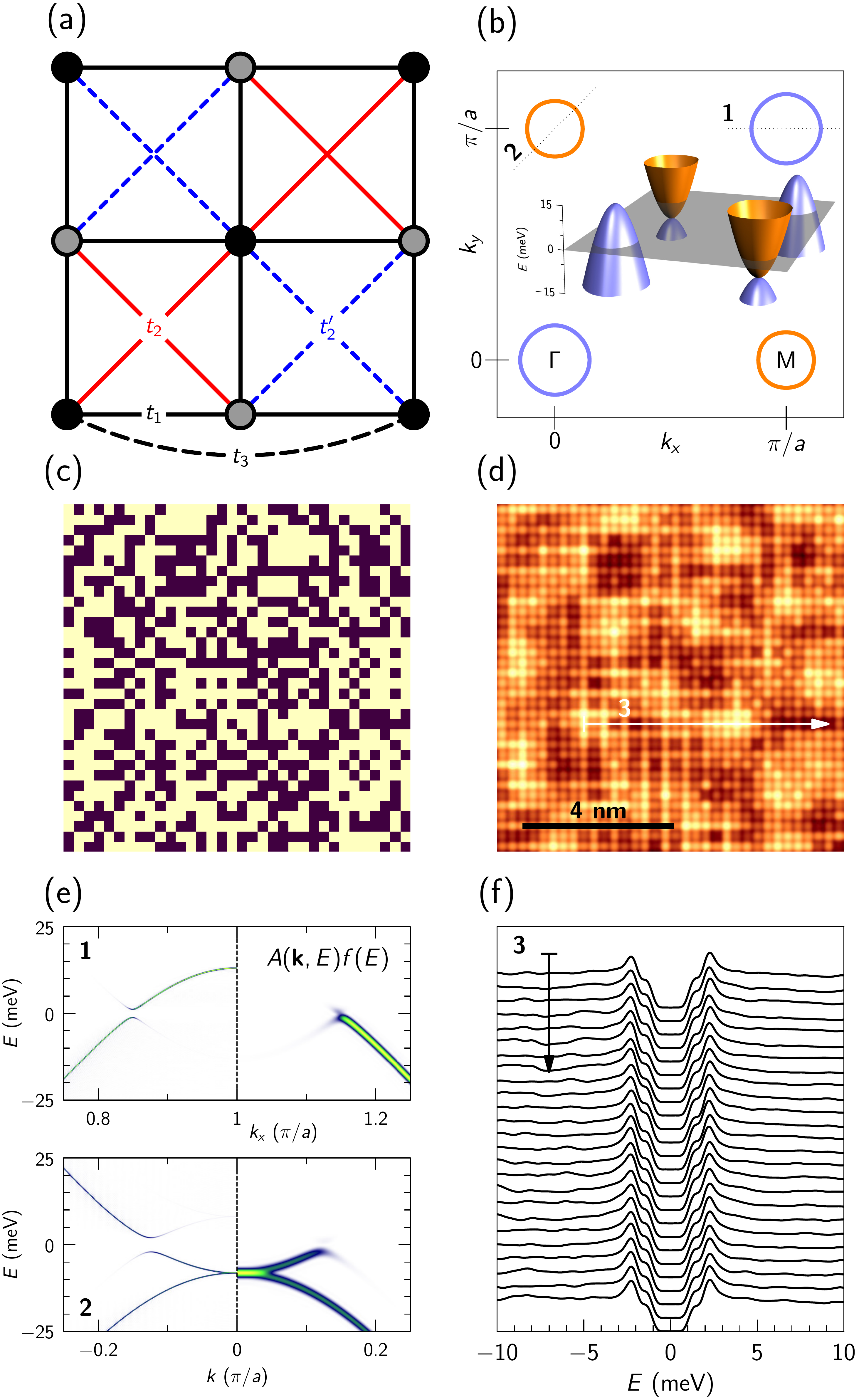}
\caption{\label{fig:fig1}
Low-energy tight-binding model for \FeTeSe{}. (a) Square lattice with two inequivalent Fe sites. The black bonds indicate a homogeneous and isotropic first-neighbor hopping amplitude $t_1$; the third-neighbor hopping $t_3$, only one of which is drawn, is also uniform and isotropic. The second-neighbor hopping amplitudes take the values $t_2$ and $t_2'$ as indicated. There is a second pair of states with the roles of $t_2$ and $t_2'$ exchanged. (b) Fermi surface in the 1-Fe Brillouin zone. The inset shows the dispersion with hole and electron bands at $\Gamma$ and M, respectively. (c) $35a\times35a$ region showing Te (bright) and Se (dark) unit cells; this represents 0.06\% of the system size used in the simulations. (d) Simulated STM tunneling topography of the same area. (e) Momentum-resolved spectral function (left part of graphs) and same quantity at lower resolution and multiplied by a Fermi function at $T=6$~K (right part of graphs) along the two cuts indicated in (b).  (f) Local density of states along the path shown in (d). The energy resolution is set to 0.64~meV ($N=7000$) in (d) and (f), and in (e) to 0.25~meV ($N=18\,000$) for the left parts and 2~meV ($N=2250$) for the right parts; see Appendix~\ref{app:resolution} for a definition of the resolution.
}
\end{figure}

The early photoemission studies have revealed a significant band renormalization; furthermore, the DFT calculations give quite different band structures close to the Fermi energy for FeSe and FeTe \cite{Subedi-2008, Tamai-2010}. These observations suggest that an effective model is preferable, in order to describe the low-energy region and the superconducting state of \FeTeSe{}, to a model based on bare DFT bands. The qualitative shape of the Fermi surface plays an important role in the quantum regime, because inhomogeneities of the condensate, in particular vortices, have dimensions comparable with the Fermi wavelength and scatter the Bogoliubov excitations of the superconductor at large angles, thus sensing the whole Fermi surface. In order to keep the approach simple while preserving the relevant ingredients for a successful low-energy theory of \FeTeSe{}, we build on the $S_4$-symmetric microscopic model introduced in Ref.~\onlinecite{Hu-2012} for iron pnictides and chalcogenides. This is a four-orbital model that captures the low-energy features of the more familiar five-orbital model \cite{Kuroki-2008, Graser-2009} and explains the robustness of the $s$-wave superconducting state in this class of materials. In the simplest variant, the model has one twofold-degenerate electron-like band and one twofold-degenerate hole-like band. The electron and hole bands follow from a tight-binding Hamiltonian with the structure shown in Fig.~\ref{fig:fig1}(a). Each Fe site is coupled to its first and third neighbors by isotropic hopping amplitudes $t_1$ and $t_3$, while the coupling to second neighbors is anisotropic and rotated by 90 degrees at the two inequivalent Fe sites. The dispersion relation measured from the chemical potential $\mu$ for these electron and hole bands is
	\begin{multline}\label{eq:xik}
		\xi^{\pm}_{\vec{k}}=4t_{2s}\cos(k_xa)\cos(k_ya)\\
		\pm2\sqrt{t_1^2[\cos(k_xa)+\cos(k_ya)]^2+[2t_{2d}\sin(k_xa)\sin(k_ya)]^2}\\
		+2t_3[\cos(2k_xa)+\cos(2k_ya)]-\mu,
	\end{multline}
where $t_{2s}=(t_2+t'_2)/2$ and $t_{2d}=(t_2-t'_2)/2$ are the $s$- and $d_{x^2-y^2}$-symmetric second-neighbor hopping amplitudes, respectively, and $a=2.69$~\AA{} is the lattice parameter of the 1-Fe unit cell. The second pair of bands has the roles of $t_2$ and $t_2'$ interchanged. The difference between $t_2$ and $t_2'$ stems from the fact that these hopping amplitudes are generated through hybridization with the out-of-plane $p$ orbitals of the pnictogen or chalcogen atom. The robust $d_{x^2-y^2}$-symmetric component $t_{2d}$ stabilizes the $s$-wave pairing on the second neighbors, very much like, in the cuprates, the $s$-symmetric first-neighbor hopping stabilizes the $d_{x^2-y^2}$ pairing symmetry \cite{Hu-2012}. For the choice of the five parameters, we require that the dispersion satisfies the following conditions: (1) the Fermi surface has a hole pocket at $\Gamma$ and an electron pocket at M; (2) the Fermi wave vectors on the $\Gamma$ and M pockets are $k_{\mathrm{F},\Gamma}=0.15\pi/a$ and $k_{\mathrm{F},\mathrm{M}}=0.12\pi/a$, respectively; (3) the hole band at $\Gamma$ has its maximum at $13$~meV; (4) the electron pocket has minimal anisotropy; (5) the self-consistent vortex-core size is 16~\AA. Condition (2) uses an average of the measured wave vectors at the $\alpha_2$ and $\alpha_3$ sheets in FeTe$_{0.42}$Se$_{0.58}$ \cite{Tamai-2010}. Condition (3) is based on the band parametrization used in Ref.~\onlinecite{Sarkar-2017}. Condition (4) is imposed for simplicity and used to fix the parameter $t_{2d}$. At low energy, the main effect of $t_{2d}$ is to change the anisotropy of the Fermi surface at the M point. One can see this by expanding the dispersion around the $\Gamma$ and M points. At leading order, the dispersion around $\Gamma$ is isotropic, while around M there is a term $t_1[(2t_{2d}/t_1)^2-1](k_x^2-k_y^2)$. The choice $t_{2d}=t_1/2$ cancels this term and minimizes the anisotropy of the M pocket. Finally, condition (5) sets the overall bandwidth, or rather the average Fermi velocity, such that the low-energy theory properly reproduces the emerging length scale that controls the vortices \cite{Shruti-2015}, as developed further below \footnote{For the determination of the bandwidth via the vortex-core size, I have mistakenly used the lattice parameter of the 2-Fe unit cell, as was realized after submission of this work. I am grateful to Lingyuan Kong for pointing this out. For this reason, the value 16~\AA{} is a factor $\sqrt{2}$ smaller than the value 22~\AA{} of the coherence length reported in Ref.~\onlinecite{Shruti-2015}. This has only minor impact on the results, because the main parameter $k_{\mathrm{F}}\xi$, being dimensionless, is not affected}. The resulting average value $k_{\mathrm{F}}\xi=2.5$ realizes the strong quantum regime to which \FeTeSe{} belongs. The model parameters satisfying conditions (1)--(5) are $(t_1,t_2,t_2',t_3,\mu)=(53,55.6,2.6,14.4,-51)$~meV. The Fermi surface and the dispersion in the low-energy sector probed by STM and angle-resolved photoemission (ARPES) experiments are depicted in Fig.~\ref{fig:fig1}(b).

To describe the superconducting state, we adopt the functional gap dependence reported in Ref.~\onlinecite{Miao-2012}. This $s^{\pm}$ state is realized in real space by means of isotropic pairing amplitudes $\Delta_2/4>0$ and $\Delta_3/4<0$ on all second- and third-neighbor bonds, respectively. The corresponding momentum-space gap structure is
	\begin{multline}
		\Delta_{\vec{k}}=\Delta_2\cos(k_xa)\cos(k_ya)\\
		+\frac{\Delta_3}{2}[\cos(2k_xa)+\cos(2k_ya)].
	\end{multline}
The sign reversal of the order parameter between the two pockets is not an essential feature for the results presented in the present study. What \emph{is} essential, though, is that the order parameter is nodeless on the Fermi surface. The ARPES values $\Delta_2=3.55$~meV and $\Delta_3=-0.95$~meV \cite{Miao-2012} produce a gap in the DOS that is approximately twice as wide as the gap seen in high-resolution STM experiments [see Fig.~\ref{fig:figA1}(b) in Appendix~\ref{app:gap}]. The optical conductivity also appears to be consistent with gaps larger than those seen by STM \cite{Homes-2010, Homes-2015}. A bump at $\sim4$~meV observed by STM \cite{Hanaguri-2010} was interpreted as the signature of a large gap \cite{Miao-2012}. However, no systematic structure at $\pm4$~meV is seen in Refs~\onlinecite{Wang-2018, Chen-2018}, despite the large number of spectra reported. This discrepancy was ascribed to momentum selectivity of the tunneling matrix element \cite{Wang-2018}, which would hide the large gap at the M points. Note that, in the cuprates, the largest gap at the M point is consistently seen with the same amplitude in photoemission and tunneling \cite{Damascelli-2003, Fischer-2007}. Since our focus here is on tunneling, we rescale the reported values by a factor of two and use $\Delta_2=1.775$~meV and $\Delta_3=-0.475$~meV. These values yield LDOS spectra in semiquantitative agreement with the STM data if the resolution of the calculation is set to 0.64~meV [compare Fig.~\ref{fig:fig1}(f) with Fig.~1c of Ref.~\onlinecite{Chen-2018}]. In particular, the robust and sharp gap edge at $\sim1$~meV, which is seen in all STM measurements \cite{Hanaguri-2010, Yin-2015, Wang-2018, Chen-2018}, is well reproduced. We also calculate the spectral function and obtain features much sharper than the ones observed in photoemission [Fig.~\ref{fig:fig1}(e)], despite averaging over a disordered region as described below. This raises the question of the origin of a broad signal in photoemission and how it may affect the determination of gap values that are similar to the resolution of the experiment (2~meV) \cite{Miao-2012}. Because the second-neighbor hopping is anisotropic, an isotropic order parameter between second neighbors can only be generated self-consistently by the Bogoliubov--de Gennes equations if the pairing interaction has a small $d_{x^2-y^2}$ component. We find that an interaction $V_2=-28.4$~meV, $V_2'=-27.0$~meV, corresponding to $V_{2s}=-27.7$~meV and $V_{2d}=-0.7$~meV, produces the desired gap structure on the second neighbors, while $V_3=-27.2$~meV generates the order parameter on the third-neighbor bonds.

Se and Te are isoelectronic, the former being slightly more electronegative than the latter. One may therefore expect the unit cells containing Se to be attractive compared with those containing Te. This disorder is weak and leads to local displacements of spectral weight and contrast in the topography measured by STM. In order to incorporate the chemical disorder, we introduce two on-site energies $V_{\mathrm{Te}}$ and $V_{\mathrm{Se}}$ and distribute them randomly with the required proportion on the Fe sites of the model [Fig.~\ref{fig:fig1}(c)]. To keep the electron density fixed, the chemical potential must be shifted by the average on-site energy $xV_{\mathrm{Te}}+(1-x)V_{\mathrm{Se}}$, where $x$ is the concentration of Te. Relative to the disorder-free chemical potential $\mu$, the on-site energies are therefore $(1-x)(V_{\mathrm{Te}}-V_{\mathrm{Se}})\equiv V$ in the unit cells containing Te and $-x/(1-x)V$ in those containing Se. This model of disorder is certainly very crude, but has the advantage of reducing the number of unknown parameters to just one. The resulting potential is random and bimodal without spatial correlations. Because, in reality, the potential on Fe atoms originates from out-of-plane disorder, a better model would attach a screened Coulomb potential to each Se and Te atom, leading to a more smooth landscape with spatial correlations at the Fe sites. Disorder models of this kind have been studied in relation to cuprate superconductors \cite{Nunner-2005, Nunner-2005b, Nunner-2006, Sulangi-2018}. This sort of refinement seems not essential for the vortex-core spectroscopy, which is the main focus of this study, and is therefore left for a future investigation. We tentatively set the strength of the potential to $V=1$~meV. In order to fix this value, we have studied the disorder-induced fluctuations of the LDOS. On each given Fe site, the surrounding disorder configuration induces in the LDOS specific structures whose amplitudes scale with $V$ (see Fig.~\ref{fig:figA2} in Appendix~\ref{app:disorder}). Simultaneously, the disorder reduces the height of the superconducting coherence peaks. With the value $V=1$~meV, the fluctuations of the calculated LDOS are comparable with the variations of the tunneling spectrum measured in zero field \cite{Chen-2018}. Figure~\ref{fig:fig1}(f) shows a series of 25 spectra taken along the path indicated in Fig.~\ref{fig:fig1}(d).

The calculations reported in this study are based on the Bogoliubov--de Gennes theory at zero temperature, solved by means of the Chebyshev-expansion technique introduced in Ref.~\onlinecite{Covaci-2010}, and using the asymmetric singular gauge of Ref.~\onlinecite{Berthod-2016} to describe disordered distributions of vortices. The Bogoliubov--de Gennes equations we consider are
	\begin{equation}\label{eq:BdG}
		\sum_{\vec{r}'}\begin{pmatrix}
		h_{\vec{r}\vec{r}'}^{\alpha}&\Delta_{\vec{r}\vec{r}'}^{\alpha}\\[0.5em]
		(\Delta_{\vec{r}'\vec{r}}^{\alpha})^* & -(h_{\vec{r}\vec{r}'}^{\alpha})^*
		\end{pmatrix}\begin{pmatrix} u_{\vec{r}'}^{\alpha} \\[0.5em] v_{\vec{r}'}^{\alpha}\end{pmatrix}
		=\varepsilon\begin{pmatrix} u_{\vec{r}}^{\alpha} \\[0.5em] v_{\vec{r}}^{\alpha}\end{pmatrix},
	\end{equation}
where $\varepsilon$ denote the energy eigenvalue for the wave function $(u_{\vec{r}}^{\alpha},v_{\vec{r}}^{\alpha})$, $\vec{r}$, $\vec{r}'$ denote the lattice sites (Fe atoms), and $\alpha$ labels the two orbitals on each Fe. $h_{\vec{r}\vec{r}'}^{\alpha}=\delta_{\vec{r}\vec{r}'}(V_{\vec{r}}-\mu)+t_{\vec{r}\vec{r}'}^{\alpha}$ contains the on-site energies $V_{\vec{r}}$ representing chemical disorder and the hopping amplitudes $t_{\vec{r}\vec{r}'}^{\alpha}$, which in zero field are as described in Fig.~\ref{fig:fig1}(a), and in a finite magnetic field carry a Peierls phase (see, e.g., Ref.~\onlinecite{Berthod-2016}). The order parameter $\Delta_{\vec{r}\vec{r}'}^{\alpha}$ is determined self-consistently from the pairing interaction $V_{\vec{r}\vec{r}'}$ according to
	\begin{equation}\label{eq:gap}
		\Delta_{\vec{r}\vec{r}'}^{\alpha}=V_{\vec{r}\vec{r}'}\sum_{\varepsilon>0}\left[
		u^{\alpha}_{\vec{r}}(v_{\vec{r}'}^{\alpha})^*f(\varepsilon)
		-(v_{\vec{r}}^{\alpha})^*u^{\alpha}_{\vec{r}'}f(-\varepsilon)\right],
	\end{equation}
where the sum runs over all eigenstates of positive energy and the Fermi factors reduce to $f(\varepsilon)=0$ and $f(-\varepsilon)=1$ at zero temperature. In order to reach our target resolution (typical separation between the energies $\varepsilon$ much smaller that the gap), we must consider square lattices of order $1500\times 1500$ sites, for which the Hamiltonian in Eq.~(\ref{eq:BdG}) has dimensions of order $8\cdot10^6\times8\cdot10^6$, preventing a direct solution by diagonalization. The Chebyshev expansion allows one to compute physical quantities like the LDOS $N(\vec{r},E)$ and the order parameter $\Delta_{\vec{r}\vec{r}'}^{\alpha}$ for each lattice point $\vec{r}$ or each bond $(\vec{r},\vec{r}')$ independently, without diagonalizing the Hamiltonian. The LDOS is computed as
	\begin{align}\label{eq:LDOS}
		N(\vec{r},E)&=-\frac{2}{\pi}\,\mathrm{Im}\sum_{\alpha}G_{\alpha\alpha}(\vec{r},\vec{r},E)\\
		\nonumber
		&\approx\frac{2}{\pi\mathfrak{a}}\mathrm{Re}\,\left\{\frac{1}{\sqrt{1-\tilde{E}^2}}\left[2+2\sum_{n=1}^N
		e^{-in\arccos(\tilde{E})}c_nK_n\right]\right\}.
	\end{align}
The first line relates the LDOS to the retarded single-particle Green's function, while the second line expresses the Chebyshev expansion, which is approximate due to truncation at finite order $N$. In the time domain, the Green's function is defined as $G_{\alpha'\alpha}(\vec{r}',\vec{r},t)=(-i/\hbar)\theta(t)\langle[\psi^{\phantom{\dagger}}_{\alpha'}(\vec{r}',t),\psi^{\dagger}_{\alpha}(\vec{r},0)]_+\rangle$, where $\theta(t)$ is the Heaviside function, $\psi^{\dagger}_{\alpha}(\vec{r})$ creates an electron at site $\vec{r}$ in orbital $\alpha$, $[\cdot,\cdot]_+$ is the anti-commutator, and the average $\langle\cdots\kern-0.03em\rangle$ is taken with respect to the Bogoliubov--de Gennes Hamiltonian $H$. The Chebyshev coefficients are
	\begin{equation}
		c_n=\sum_{\alpha}\langle\vec{r}\alpha|T_n(\tilde{H})|\vec{r}\alpha\rangle,
	\end{equation}
where $|\vec{r}\alpha\rangle$ is the state representing an electron localized at point $\vec{r}$ in orbital $\alpha$ and $T_n(\tilde{H})$ is the Chebyshev polynomial of order $n$ evaluated at the dimensionless rescaled Hamiltonian $\tilde{H}=(H-\mathfrak{b})/\mathfrak{a}$. Likewise, $\tilde{E}=(E-\mathfrak{b})/\mathfrak{a}$ with, in our case, $\mathfrak{a}=605$~meV and $\mathfrak{b}=162$~meV. The coefficients $K_n=\{(N-n+1)\cos[\pi n/(N+1)]+\sin[\pi n/(N+1)]\cot[\pi/(N+1)]\}/(N+1)$ remove unphysical oscillations due to the truncation at order $N$ \cite{Weisse-2006}. All calculations are performed on a lattice comprising 2\,002\,001 sites centered on the site where the LDOS is being calculated \cite{Berthod-2016}. The calculation scales with the order $N$, the energy resolution improving like $1/N$, as illustrated in Appendix~\ref{app:resolution}. The simulated topography of Fig.~\ref{fig:fig1}(d) was constructed by calculating the tunneling current at each of the $35\times35$ sites shown in Fig.~\ref{fig:fig1}(c) as $I(\vec{r})=\int_0^{10~\mathrm{meV}}dE\,N(\vec{r},E)$, consistently with the experimental protocol \cite{Chen-2018}, and attaching to each pixel a function with the shape of a rounded square colored by $I(\vec{r})$. The spatial contrast results from LDOS fluctuations due to disorder. Notice that the spatial tunneling current distribution is not bimodal and fails to show one-to-one correspondence with the distribution of Te and Se potentials, despite clear correlations between Figs.~\ref{fig:fig1}(c) and \ref{fig:fig1}(d). We find a weak positive correlation coefficient $R=0.4$ (see Appendix~\ref{app:correlation}). The simulated topography shows dark patches (lower current) in Se-rich regions and bright patches in Te-rich regions, in agreement with the experimental observations \cite{Kato-2009, Tamai-2010, Hanaguri-2010, He-2011, Lin-2013, Yin-2015, Wang-2018, Chen-2018}. A trace of 25 LDOS spectra is displayed in Fig.~\ref{fig:fig1}(f). The smaller and larger gaps, residing on the $\Gamma$ and M pockets, respectively, are visible despite the relatively low resolution used (see Appendix~\ref{app:gap} for high-resolution LDOS curves). The disorder induces fluctuations in $N(\vec{r},E)$, in particular fluctuations of the coherence-peak height. One also notices small fluctuations of the gap width as reported in Ref.~\onlinecite{Lin-2013}. Those are mostly due to the disorder, but also reflect the self-consistent adjustment of the order parameter in the disordered landscape.

A few remarks are in order regarding self-consistency. The gap equation (\ref{eq:gap}) can be rewritten in a form similar to Eq.~(\ref{eq:LDOS}) involving the anomalous Green's function and the corresponding Chebyshev coefficients \cite{Berthod-2016}. In principle, Eq.~(\ref{eq:gap}) must be solved self-consistently on each bond ($\vec{r},\vec{r}'$). In a system of two millions sites, this is presently beyond reach. For calculating the LDOS at a given site, however, precise self-consistent values of the order parameter at distant sites are not required and can be replaced by approximate values. For the data shown in Figs.~\ref{fig:fig1}(d)--\ref{fig:fig1}(f), we limited the search for a self-consistent solution to the region shown in Fig.~\ref{fig:fig1}(c), including a few sites around it, while keeping the order parameter fixed to its unperturbed value in the rest of the system. The self-consistency brings only small quantitative changes to the various properties. In particular, the spectral gap defined by half the energy separation between the two main coherence peaks differs by less than 0.2\% on average (0.5\% at maximum) between the self-consistent and non-self-consistent calculations (see Fig.~\ref{fig:figA3} in Appendix~\ref{app:correlation}).

The real-space Green's function also gives access to the momentum-resolved spectral function measured by ARPES. For a translation-invariant system, the Green's function depends on the relative coordinate $\vec{r}'-\vec{r}$ and the spectral function $A(\vec{k},E)$ is proportional to the imaginary part of its Fourier transform with respect to $\vec{r}'-\vec{r}$. In a disordered system like \FeTeSe{}, the spectral function must be spatially averaged over the center-of-mass coordinate $(\vec{r}'+\vec{r})/2$, leading us to the expression
	\begin{align}\label{eq:ARPES}
		A(\vec{k},E)&=\frac{1}{M}\sum_{\vec{r}}\left({-\frac{2}{\pi}}\right)
		\mathrm{Im}\sum_{\vec{r}'}e^{-i\vec{k}\cdot(\vec{r}'-\vec{r})}\sum_{\alpha\alpha'}
		G_{\alpha'\alpha}(\vec{r}',\vec{r},E)\\
		\nonumber
		&\approx\frac{2}{\pi\mathfrak{a}}\mathrm{Re}\,\left\{\frac{1}{\sqrt{1-\tilde{E}^2}}
		\left[2+2\sum_{n=1}^Ne^{-in\arccos(\tilde{E})}c'_nK_n^{\phantom{'}}\right]\right\}.
	\end{align}
The spatial average is performed on $M$ sites labeled $\vec{r}$, while the $\vec{r}'$ sum runs over the whole lattice. The Chebyshev expansion has the same form as for the LDOS, except that the coefficients now depend on nonlocal matrix elements according to
	\begin{equation}
		c'_n=\frac{1}{M}\sum_{\vec{r}}\sum_{\alpha\alpha'}\sum_{\vec{r}'}
		e^{-i\vec{k}\cdot(\vec{r}'-\vec{r})}\langle\vec{r}'\alpha'|T_n(\tilde{H})|\vec{r}\alpha\rangle.
	\end{equation}
Figure~\ref{fig:fig1}(e) shows $A(\vec{k},E)$ evaluated along the two cuts drawn in Fig.~\ref{fig:fig1}(b), as well as $A(\vec{k},E)f(E)$ calculated with a lower resolution for a temperature $T=6$~K [this temperature applies only to $f(E)$; $A(\vec{k},E)$ remains the zero-temperature result]. Since the calculation of the spectral function is relatively time-consuming, the spatial average was restricted to the region shown in Figs.~\ref{fig:fig1}(c) and \ref{fig:fig1}(d). Notice that the spatial average is not a disorder average in the usual meaning of statistical average over the disorder: here, a single configuration of the disorder was generated and kept for the whole study. It is seen that the disorder has little effect on $A(\vec{k},E)$. In particular, no broadening is observed: $A(\vec{k},E)$ has the resolution implied by the Chebyshev expansion. Note that the broadening of order $\pi V^2N(0)$ that would be expected from a conventional disorder average, where $N(0)$ is the Fermi-level DOS, is here very small ($\sim 0.04$~meV). The gaps on the hole and electron pockets are visible, even if the resolution is lowered to 2~meV. References~\onlinecite{Tamai-2010, Miao-2012} report different responses of the hole bands around $\Gamma$ to light polarization, attributed to their different orbital characters. These are optical selection rules associated with the photoemission matrix element. In the present model, the spectral function itself has structure in momentum space, independent of any matrix-element effect, and, interestingly, the hole band has exactly zero spectral weight along $\Gamma$--M (see Appendix~\ref{app:ARPES}). For this reason, the hole band was imaged around $(\pi,\pi)$ in Fig.~\ref{fig:fig1}(e).

\section{Isolated vortex without chemical disorder}
\label{sec:vortex}

\begin{figure}[tb]
\includegraphics[width=0.8\columnwidth]{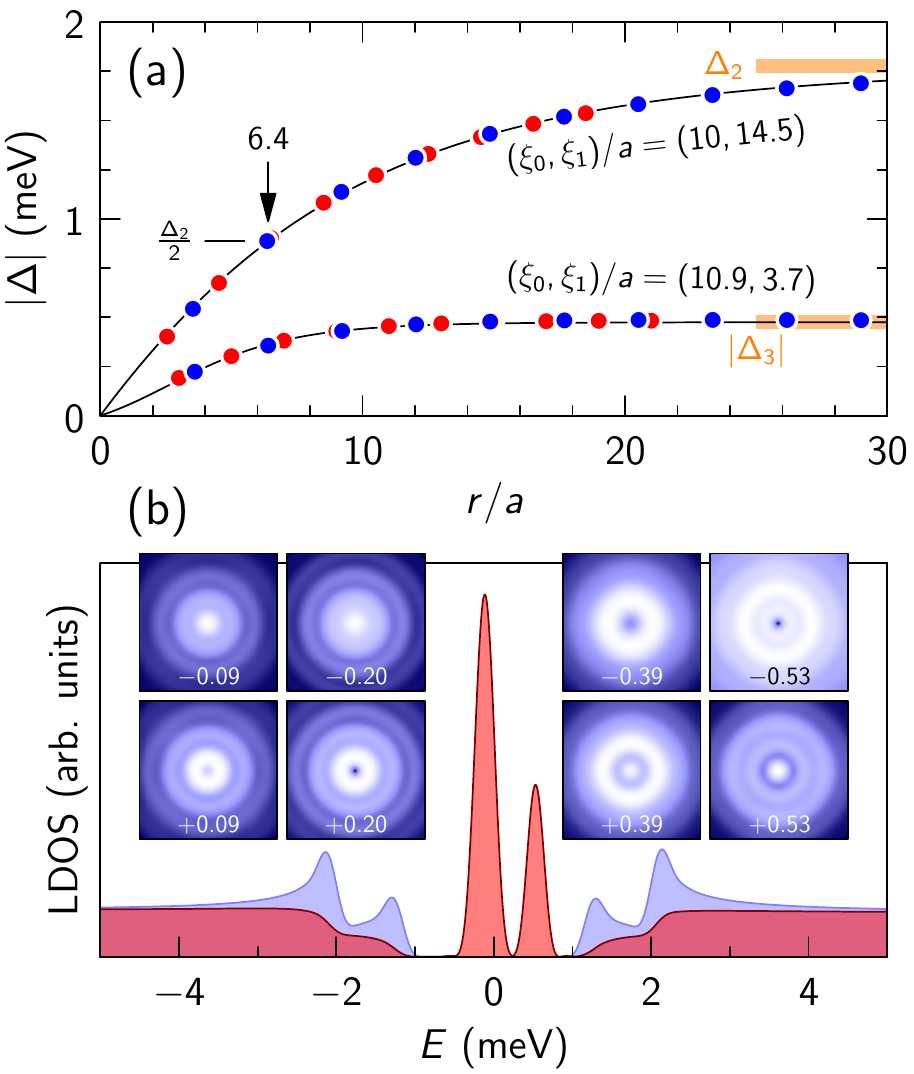}
\caption{\label{fig:fig2}
(a) Self-consistent vortex order parameter. The red and blue points show the order-parameter modulus (multiplied by four) on bonds connected to lattice sites along the (10) and (11) direction, respectively; for instance, the two red points closest to the vortex center correspond to lattice coordinates (2,0), the center of the bond being at (2.5,0.5) for the large gap and (3,0) for the small gap. The solid lines show the interpolation formula given in the text. Note the absence of anisotropy between the (10) and (11) directions. (b) LDOS at the vortex center (red) and without vortex (blue) calculated with a resolution of 0.25~meV ($N=18\,000$). The insets show the spatial distribution of the LDOS in a $51a\times51a$ region around the vortex at the energies of the low-lying bound states. The color scale is logarithmic, going from blue (minimal intensity) to white (maximal intensity).
}
\end{figure}

We turn now to the vortex-core states, starting with a single isolated vortex without chemical disorder. The effects of disorder and nonideal vortex lattice are the topics of the next section. The order parameter $\Delta_{\vec{r}\vec{r}'}^{\alpha}$ for a vortex has a modulus that decreases from its asymptotic bulk value to zero when $\vec{r}$ and $\vec{r}'$ approach the vortex center, and a phase that winds by $2\pi$ along any path encircling the vortex. Solving Eq.~(\ref{eq:gap}) in the neighborhood of the vortex, we find that the self-consistent order parameter modulus is very nearly isotropic around the vortex center and accurately parametrized by the form \cite{Berthod-2016} $\Delta(r)=\Delta_0/[1+(\xi_0/r)\exp(-r/\xi_1)]$, where $r$ is the distance from the vortex center to the center of the bond where the order parameter $\Delta$ is located, $\Delta_0$ is the asymptotic bulk value (either $\Delta_2$ or $\Delta_3$), and $\xi_0$, $\xi_1$ are two parameters with the unit of length. This analytical expression is compared with the self-consistent profile in Fig.~\ref{fig:fig2}(a). The core sizes $\xi_c$, defined by $\Delta(\xi_c)=\Delta_0/2$, equal $6.4a$ for the large gap and $3.85a$ for the small gap, which, once weighted by the bulk gap amplitudes and averaged, give the overall core size of 16~\AA. Recall that the bandwidth was adjusted such as to achieve this value for the core size \cite{Note1}. Note also that no unique definition of the core size exists, and that the value obtained with our definition based solely on the self-consistent order parameter may differ from values deduced from, e.g., spectroscopic data. Regarding the phase of the vortex order parameter, we find that, within numerical accuracy, it is given by the angle defined by the center of the bond and the center of the vortex.

\begin{figure}[tb]
\includegraphics[width=\columnwidth]{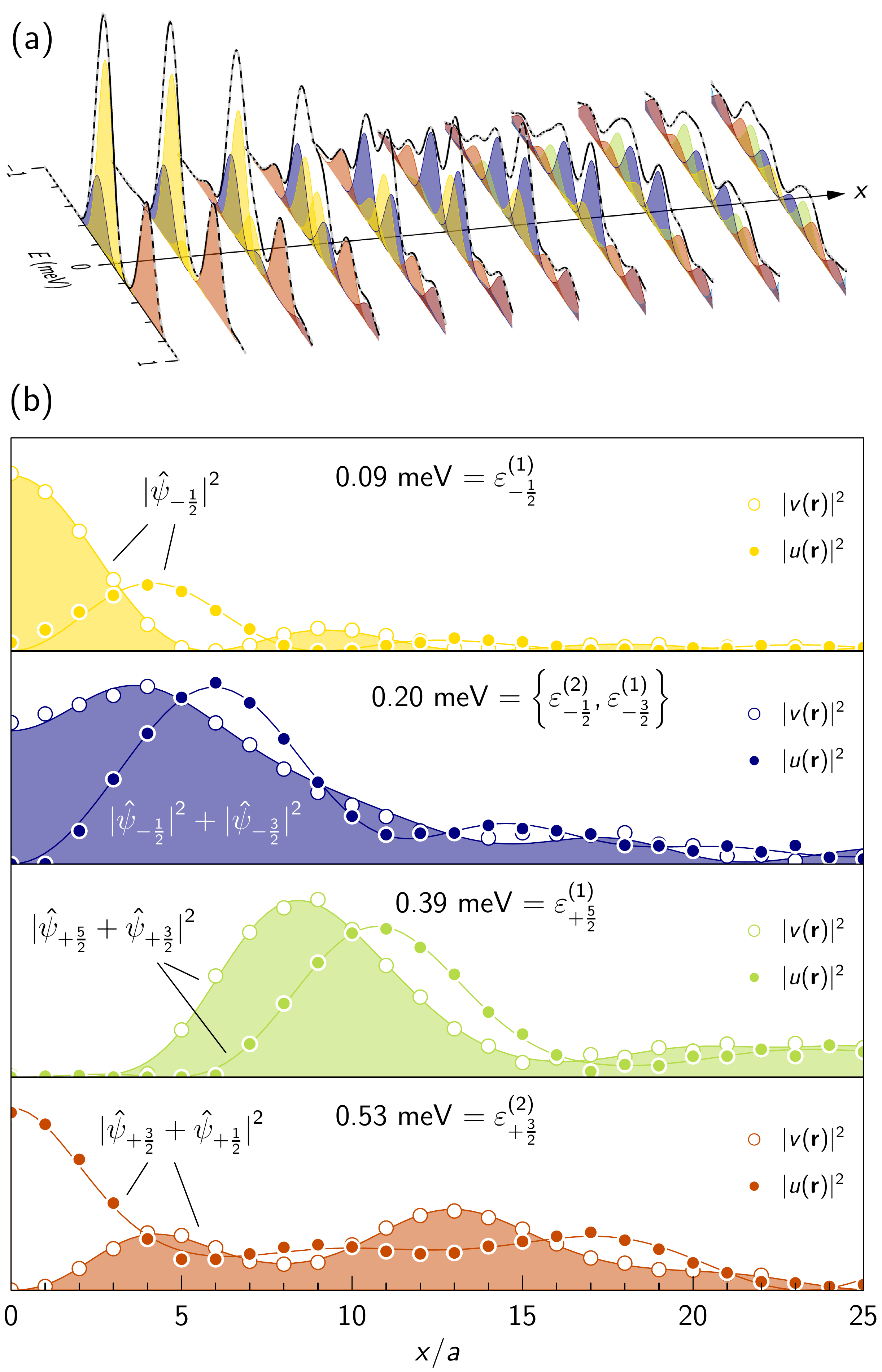}
\caption{\label{fig:fig3}
(a) Spectral decomposition of the vortex LDOS. The black curves show the LDOS at sites $(ia,0)$ for $i=0$ (left) to $i=10$ (right), the shaded curves show the pairs of peaks $|u_{\varepsilon}(\vec{r})|^2\delta_N(E-\varepsilon)$ and  $|v_{\varepsilon}(\vec{r})|^2\delta_N(E+\varepsilon)$ (one color per pair), and the dashed gray curves show the sum of all these peaks. (b) Amplitudes $|u_{\varepsilon}(\vec{r})|^2$ (full symbols) and $|v_{\varepsilon}(\vec{r})|^2$ (empty symbols) for the four lowest-energy states and their interpretation in terms of angular-momentum eigenstates (solid lines).
}
\end{figure}

The LDOS at the vortex center shown in Fig.~\ref{fig:fig2}(b) presents two peaks near $-0.11$~meV and $+0.53$~meV, reminiscent of the Caroli--de Gennes--Matricon bound states \cite{Caroli-1964}. The width and shape of the peak at $+0.53$~meV are identical to those of the resolution function (see Appendix~\ref{app:resolution}), indicating that it corresponds to a single bound state. This calculation was performed with $N=18\,000$, achieving an energy resolution of 0.25~meV. For the taller peak at $-0.11$~meV, the width is slightly larger than the resolution: a spectral analysis reveals that this peak gets contributions from two bound states. As our calculation method delivers the LDOS in terms of the Green's function, Eq.~(\ref{eq:LDOS}), we miss a direct access to the energies and wave functions of the Bogoliubov--de Gennes Hamiltonian. Nonetheless, in the subgap range where the spectrum is discrete, it is possible to extract the energies and wave functions by fitting the expression
	\begin{equation}\label{eq:LDOS2}
		N(\vec{r},E)=\sum_{\varepsilon}\left[|u_{\varepsilon}(\vec{r})|^2\delta_N(E-\varepsilon)
		+|v_{\varepsilon}(\vec{r})|^2\delta_N(E+\varepsilon)\right]
	\end{equation}
to the calculated LDOS. $u_{\varepsilon}(\vec{r})$ and $v_{\varepsilon}(\vec{r})$ are the electron and hole amplitudes, respectively, and $\varepsilon>0$ are the energies. Equation~(\ref{eq:LDOS2}) gives the exact LDOS if the energies $\varepsilon$ are allowed to run over all states, both in the continuous and discrete parts of the spectrum, and if $\delta_N(E)$ is the Dirac delta function. In our case, we restrict ourselves to a small set of discrete energies in the subgap region and we replace the Dirac delta function by the resolution function of the Chebyshev expansion. We consider a series of 26 LDOS spectra taken along the (10) direction at the positions $(x,y)=(ia,0)$, $i=0,\ldots,25$, and we fit Eq.~(\ref{eq:LDOS2}) to this whole data set at once. The outcome of this fitting is displayed in Fig.~\ref{fig:fig3}. We expect the result to be independent of the direction because, as can be seen in the insets of Fig.~\ref{fig:fig2}(b), the LDOS is almost perfectly isotropic. Figure~\ref{fig:fig3}(a) shows the spectral decomposition of the LDOS as a sum of resolution-limited peaks. Only the LDOS curves up to $x=10a$ are shown for clarity; the complete data set is displayed in Fig.~\ref{fig:figA4}, Appendix~\ref{app:decomposition}. The quality of the fit is very good, although not perfect. Given the large number of adjustable parameters, finding the absolute minimum is somewhat chancy. Nevertheless, the results make sense and are sufficient for our purposes.

\begin{figure}[tb]
\includegraphics[width=0.8\columnwidth]{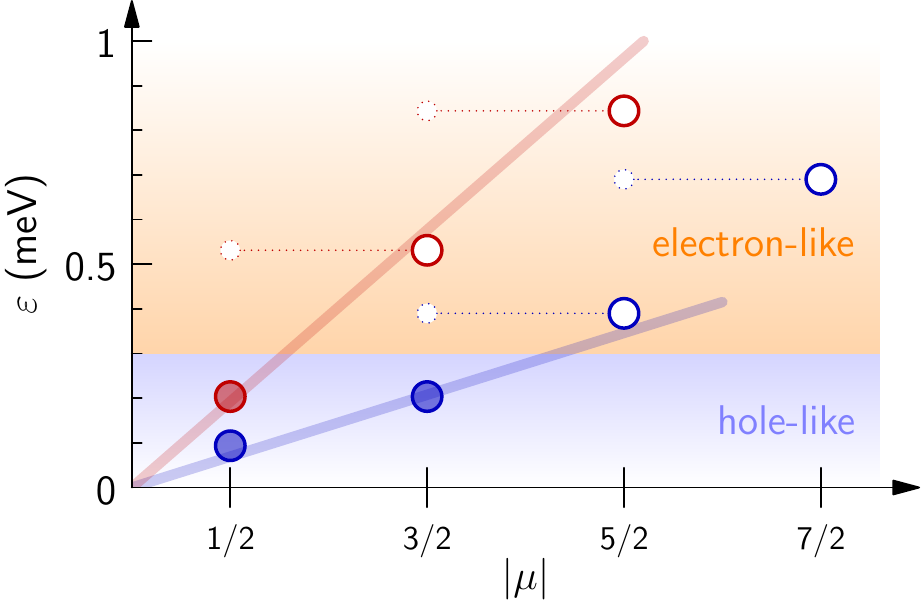}
\caption{\label{fig:fig4}
Spectrum of subgap bound states for an isolated vortex, classified according to their approximate angular momentum $\mu$. Filled (empty) symbols correspond to hole-like (electron-like) states with negative (positive) $\mu$. Dashed lines indicate mixing of angular momenta.The blue and red straight lines show the behavior $|\mu|\Delta^2/E_{\mathrm{F}}$ with $\Delta=1.2$ and 2~meV, respectively, and $E_{\mathrm{F}}=10.4$~meV.
}
\end{figure}

In the continuum model, an isolated vortex has cylindrical symmetry and the eigenstates can be chosen such that the electron and hole components $u$ and $v$ have well-defined angular momenta \cite{Caroli-1964, Bardeen-1969, Gygi-1990a, Hayashi-1998}. The electron component has angular momentum $\mu-1/2$ and behaves in the core like $J_{\mu-1/2}(k_+r)$, where $\mu$ is half an odd integer and $k_+$ is a wave vector slightly above $k_{\mathrm{F}}$, while the hole component has angular momentum $-\mu-1/2$ and behaves like $J_{\mu+1/2}(k_-r)$. We conform to the traditional notation for the angular momentum $\mu$, since confusion with the chemical potential is unlikely. All Bessel functions $J_n$ except $J_0$ vanish at the origin, such that the only angular-momentum eigenstates with a finite amplitude at the vortex center are $\mu=\pm 1/2$. For an electron band (i.e., a positive mass), the lowest Bogoliubov excitation is electron-like with momentum $+1/2$ and finite value of $u$ at the vortex center. For a hole band, the lowest excitation is hole-like with momentum $-1/2$ and finite value of $v$ in the core \cite{Araujo-2009}. The angular momentum is not a good quantum number on the lattice. Nevertheless, because the Fermi energy is small in our model, the dispersion is nearly parabolic [see Fig.~\ref{fig:fig1}(b)] and the effective low-energy Hamiltonian resembles the continuum, as exemplified by the approximate cylindrical symmetry of the LDOS in Fig.~\ref{fig:fig2}(b). One sees in Fig.~\ref{fig:fig2}(b) that the lowest bound state ($\varepsilon=0.09$~meV) has its hole component finite in the core ($\propto J_0$), while the electron component vanishes ($\propto J_1$) [see also the top panel in Fig.~\ref{fig:fig3}(b)]. This suggests that the low-energy carriers are holes. At first sight, one expects that the next bound state at $\varepsilon=0.2$~meV has angular momentum $-3/2$ with its hole (electron) component vanishing in the core like $J_1$ ($J_2$), an expectation contradicted by the data in Figs.~\ref{fig:fig2}(b) and \ref{fig:fig3}(b). Here we must remember that the model has four bands grouped in two degenerate pairs. The correct expectation is therefore to observe two doubly-degenerate states at each angular momentum. The second state at $\varepsilon=0.2$~meV indeed has angular momentum $-1/2$ and does not vanish in the core. In order to determine the angular momenta of the various bound states, we fit the form
	\begin{equation}\label{eq:Bessel}
		\hat{\psi}_{\mu}(\vec{r})=\begin{pmatrix}u^{(1)}J_{\mu-\frac{1}{2}}\big(k_+^{(1)}r\big)
		+u^{(2)}J_{\mu-\frac{1}{2}}\big(k_+^{(2)}r\big)\\[1em]
		v^{(1)}J_{\mu+\frac{1}{2}}\big(k_-^{(1)}r\big)
		+v^{(2)}J_{\mu+\frac{1}{2}}\big(k_-^{(2)}r\big)\end{pmatrix}
	\end{equation}
to the electron and hole components extracted from the LDOS via Eq.~(\ref{eq:LDOS2}). Equation~(\ref{eq:Bessel}) represents an angular-momentum eigenstate with radial components at two different wave vectors, as appropriate in a two-band setup. The state at $\varepsilon=0.09$~meV is very well approximated by $\hat{\psi}_{-1/2}$, confirming that this is a hole-like state with momentum $-1/2$ [see Fig.~\ref{fig:fig3}(b)]. At $\varepsilon=0.2$~meV, the fit yields two independent components with angular momenta $-1/2$ and $-3/2$. We believe that two bound states live there, too close in energy to be resolved. One is the second hole-like state with momentum $-1/2$, the other is the first state with momentum $-3/2$. The near degeneracy of two states here will be confirmed below (Fig.~\ref{fig:fig6}). Something quite interesting happens as the energy increases: we find that the best fit to the state at $\varepsilon=0.39$~meV is $\hat{\psi}_{+5/2}$ with a small admixture of $\hat{\psi}_{+3/2}$. Hence this is an \emph{electron}-like state and the vortex electronic structure changes from hole- to electron-like between 0.2 and 0.4~meV. This statement is confirmed by the state at $\varepsilon=0.53$~meV, as well as two other states not shown in Fig.~\ref{fig:fig3}(b) at $0.69$ and $0.84$~meV. The superposition of two angular momenta signals the broken rotational symmetry, and/or the interband mixing. At $\varepsilon=0.53$~meV, we find a $+3/2$ state with a significant mixing of $+1/2$, which explains why this state has amplitude at the vortex center despite its relatively high energy. In Ref.~\onlinecite{Kaneko-2012}, a similar mixing was advocated to explain a second peak in YNi$_2$B$_2$C. Finally, states at 0.69 and 0.84~meV are found to have angular momenta $+7/2$ and $+5/2$ with admixture of $+5/2$ and $+3/2$, respectively. The data are summarized in Fig.~\ref{fig:fig4}. The two series of bound states obey approximately the scaling $\varepsilon=|\mu|\Delta^2/E_{\mathrm{F}}$ if we take for $\Delta$ the two spectral gaps of 1.2 and 2~meV, and for $E_{\mathrm{F}}$ the value 10.4~meV, which is the average of the hole-band maximum at 13~meV and the electron-band minimum at $-7.8$~meV [see Fig.~\ref{fig:fig1}(b)].

\section{Effects of chemical disorder, density inhomogeneity, and neighboring vortices on the vortex-core spectrum}
\label{sec:disorder}

\begin{figure}[tb]
\includegraphics[width=\columnwidth]{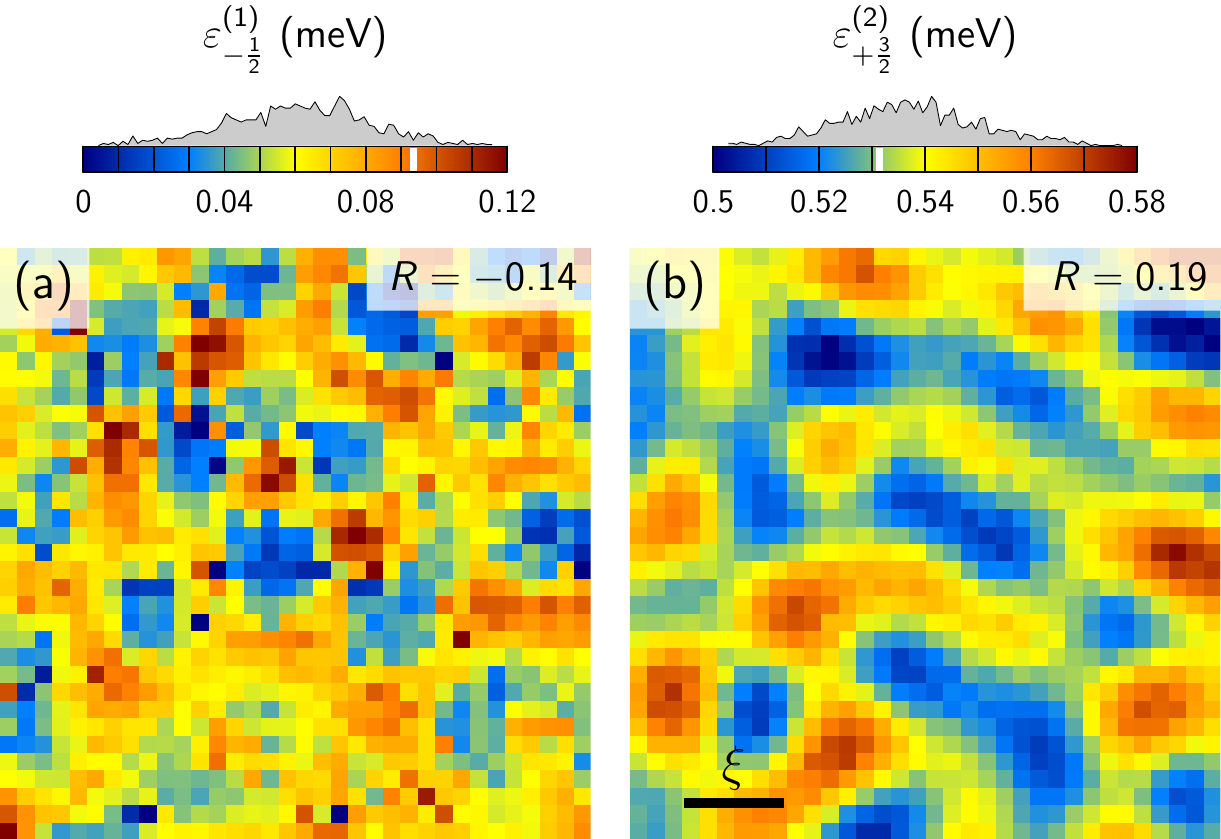}
\caption{\label{fig:fig5}
Energy of the vortex bound state $\varepsilon_{-1/2}^{(1)}$ (a) and $\varepsilon_{+3/2}^{(2)}$ (b) as the vortex is moved over the disorder. Each pixel represents a different calculation with the vortex centered at that pixel. The correlation coefficients $R$ indicate weak negative and positive correlation with the disorder. The color scales, together with the histograms of energies, are shown on top of each graph. The white bars indicate the bound-state energy in the absence of disorder. The scale bar in (b) shows the coherence length (average vortex-core size).
}
\end{figure}

Strong local scattering centers induce low-energy states in superconductors \cite{Balatsky-2006}. The effects of weak extended disorder like the chemical disorder in \FeTeSe{} have been much less studied (for a recent perspective, see Ref.~\onlinecite{Sulangi-2018} and references therein). As the experiments \cite{Chen-2018} report variations in the spectroscopy from one vortex to the next, we first ask whether these variations could be attributed to the chemical disorder. For this purpose, we consider an isolated vortex and move it across the disordered landscape of Fig.~\ref{fig:fig1}(c). For the order parameter, we use the approximate form deduced from the self-consistent calculation and shown in Fig.~\ref{fig:fig2}(a). As the self-consistent adjustment of the order parameter in the disorder has almost no effect on the LDOS (Appendix~\ref{app:correlation}), we neglect it here. For each of the $35\times35$ possible positions, we compute the LDOS at the vortex center. Overall, we find that the changes are tiny relative to the spectrum in the clean case [Fig.~\ref{fig:fig2}(b)]: all LDOS curves have the same general structure with one tall peak at small negative energy and one weaker peak at higher positive energy. We have previously attributed the tall peak to the superposition of two hole-like bound states of angular momentum $-1/2$ and energies 0.09 and 0.2~meV. The weaker peak belongs to a single electron-like state at energy 0.53~meV with principal angular momentum $+3/2$ and an admixture of $+1/2$. We fit Eq.~(\ref{eq:LDOS2}) to each LDOS curve in order to determine the correlation between bound states and disorder. The energies of the bound states vary as the vortex is moved over the disorder as shown in Fig.~\ref{fig:fig5}. The distribution of energies for each bound state is much narrower (typical variance 0.02~meV) than the corrugation of the potential (1~meV). The lowest state $\varepsilon_{-1/2}^{(1)}$, being the most localized [see Fig.~\ref{fig:fig3}(b)], varies on the same scale as the disorder and shows a weak anti-correlation with it: there is a tendency for the state to have higher energy in attractive Se-rich regions and lower energy in repulsive Te-rich regions. These tendencies do not compensate, though, and on average the energy $\varepsilon_{-1/2}^{(1)}$ is pushed down ($0.06$~meV) compared with the clean case ($0.09$~meV). This average shift occurs in spite of the fact that, on average, the potential due to disorder is zero. The behavior of the state $\varepsilon_{-1/2}^{(2)}$ is similar (not shown in the figure), except that it displays almost no shift on average. This may be due to the fact that this state is closer to the particle-hole crossover energy. The less localized state $\varepsilon_{+3/2}^{(2)}$ varies on a coarse-grained scale of the order of the coherence length [Fig.~\ref{fig:fig5}(b)]. This state has a weak \emph{positive} correlation with the disorder and is pushed on average \emph{up}. Hence electron-like and hole like vortex states behave differently in the presence of extended disorder.

\begin{figure}[tb]
\includegraphics[width=0.7\columnwidth]{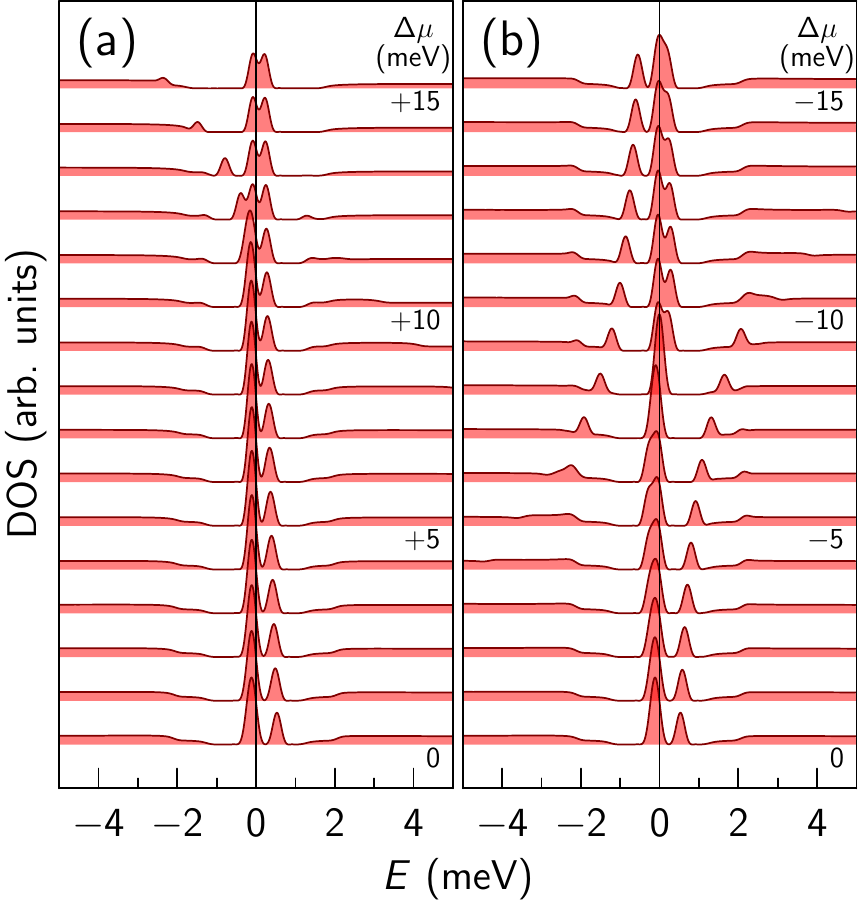}
\caption{\label{fig:fig6}
Evolution of the LDOS at the vortex center with varying the chemical potential, all other parameters held fixed, in the absence of disorder. (a) and (b) show data for increasing charge accumulation and charge depletion, respectively, from bottom to top. The curves are shifted vertically for clarity. The resolution is 0.25~meV ($N=18\,000$).
}
\end{figure}

Within our model, the chemical disorder can not explain the variety of vortex-core spectra observed in \FeTeSe{}, with sometimes a single peak, sometimes two peaks of equal heights, and sometimes a taller peak at negative energy, like in Fig.~\ref{fig:fig2}(b). As a second possibility, we study how the vortex spectrum changes when varying the chemical potential, as would occur in the presence of fluctuations in the electronic density on length scales comparable with or larger than the coherence length. It is not obvious how such nanoscale variations could be stabilized. The vortex charging effect \cite{Khomskii-1995, Kolacek-2001} is small and may explain a depletion of order $10^{-5}$ electrons per cell in the vortex, corresponding to a tiny change of chemical potential in the $\mu$eV range. Much larger changes are needed in order to produce appreciable variations in the vortex spectrum. Figure~\ref{fig:fig6} shows the evolution of the spectrum at the vortex center upon charge accumulation [Fig.~\ref{fig:fig6}(a)] and charge depletion [Fig.~\ref{fig:fig6}(b)] stemming from variations of $\mu$ by $\pm15$~meV, which correspond roughly to $\pm 0.2$ electrons per cell. The vortex order parameter was held fixed in this calculation and only $\mu$ was varied. It is seen that the charge accumulation has little effect, until $\Delta\mu=13$~meV, which is the point where the hole pocket at the $\Gamma$ point disappears [see Fig.~\ref{fig:fig1}(b)]. From this point on, one of the two hole-like states forming the tall peak at negative energy moves towards the gap edge, leaving a single pair of peaks corresponding to the states $\varepsilon_{-1/2}^{(1)}$ and $\varepsilon_{+3/2}^{(2)}$ in the vortex core. The charge depletion alters the spectrum more dramatically. As the chemical potential is lowered and the electron band at the M point is progressively emptied, the electron-like state $\varepsilon_{+3/2}^{(2)}$ moves to the gap edge and another hole-like state comes in from the negative-energy gap edge. The two hole-like states $\varepsilon_{-1/2}^{(1)}$ and $\varepsilon_{-1/2}^{(2)}$ show a very intriguing dependence on $\mu$: they initially begin to split for low values of $\Delta\mu$, before merging again into a single peak, splitting across zero energy, and merging back a second time. Most remarkably, the first merging leads to a peak at exactly zero energy: this occurs for $\Delta\mu=-7.8$~meV, which puts the chemical potential at the quadratic touching point where the electron pocket at M disappears and a hole pocket appears instead [Fig.~\ref{fig:fig1}(b)]. This accidental zero-energy peak is unrelated to Majorana-type physics \cite{Zhang-2018, Wang-2018}. Although Fig.~\ref{fig:fig6} presents some spectral variation, we believe that the source of the variability in the experimental spectra must be searched for elsewhere. In addition to being unlikely for electrostatic reasons, the large variations of chemical potential do not produce the kind of spectra that are seen by STM.

\begin{figure}[tb]
\includegraphics[width=0.7\columnwidth]{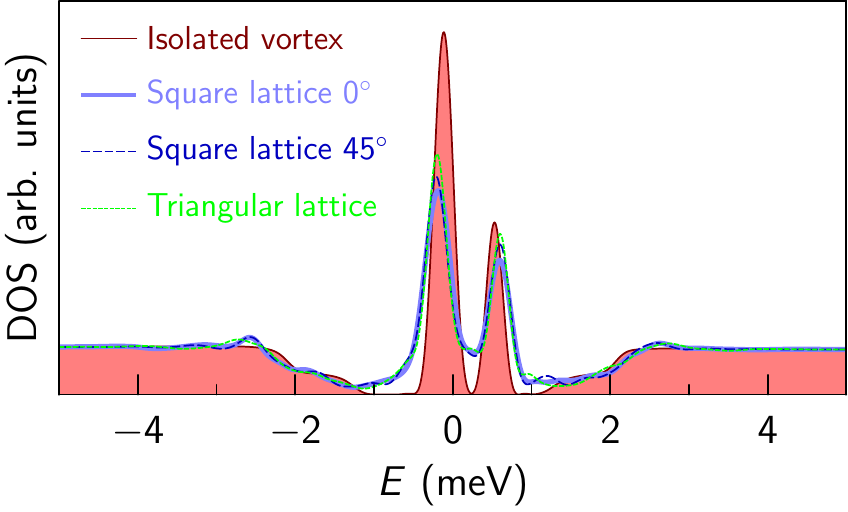}
\caption{\label{fig:fig7}
Comparison of the vortex-core LDOS in an isolated vortex (red) and in three vortex lattices for a 5~T field. The square vortex lattice has the same orientation as the Fe lattice ($0^{\circ}$), or is rotated by 45 degrees. The triangular lattice has its main axis along the Fe lattice. The resolution is 0.25~meV ($N=18\,000$).
}
\end{figure}

Finally, we investigate the influence of nearby vortices. In a perfect vortex lattice, the localized states of the isolated vortices hybridize to form bands. In a 5~T field, the hybridization is weaker than the typical energy separation between core states and the latter remain visible in the LDOS as well-defined peaks. Apart from an overall broadening, no significant qualitative change is expected in a field if the vortices are ordered. Figure~\ref{fig:fig7} shows the results for square and triangular vortex lattices of various orientations. The tiny dependence on vortex-lattice structure and orientation reflects the absence of anisotropy in the isolated vortex (Fig.~\ref{fig:fig2}), owing to low electron density and exponential localization of the bound states in the cores. This contrasts with the case of $d$-wave superconductors, where the isolated vortex defines preferred directions relative to the microscopic lattice \cite{Yasui-1999, Han-2002, Berthod-2016, Berthod-2017}. Note that the vortex-lattice order parameter was constructed by superimposing the modulus and winding phase of isolated vortices using the method described in Ref.~\onlinecite{Berthod-2016}, and a small self-consistent adjustment was neglected. We expect this to play no role for the LDOS inside the cores. The calculation is made in the limit of a large penetration depth, assuming a constant field throughout space. The measurements of Ref.~\onlinecite{Chen-2018} show a vortex distribution that is largely random at 5~T. The positional disorder in the vortices may explain the variety of spectra observed, as we show now. We have extracted vortex positions from Fig.~1d of Ref.~\onlinecite{Chen-2018}. As the simulation uses a system $\sim 4$ times larger than the field of view of that figure, we have generated random vortex positions outside the field of view with a distribution similar to that seen inside (Appendix~\ref{app:vortices}). The disordered vortices are surrounded by an ordered square lattice at the same field in order to ensure the correct boundary condition for an infinite distribution of vortices \cite{Berthod-2016}. The unknown vortex positions outside the field of view do influence the LDOS calculated inside \cite{Berthod-2017}. Hence the simulation should not be seen as an attempt to precisely model the experimental data, but as a way of obtaining typical spectra that may occur in a disordered vortex configuration.
\begin{figure}[tb]
\includegraphics[width=0.7\columnwidth]{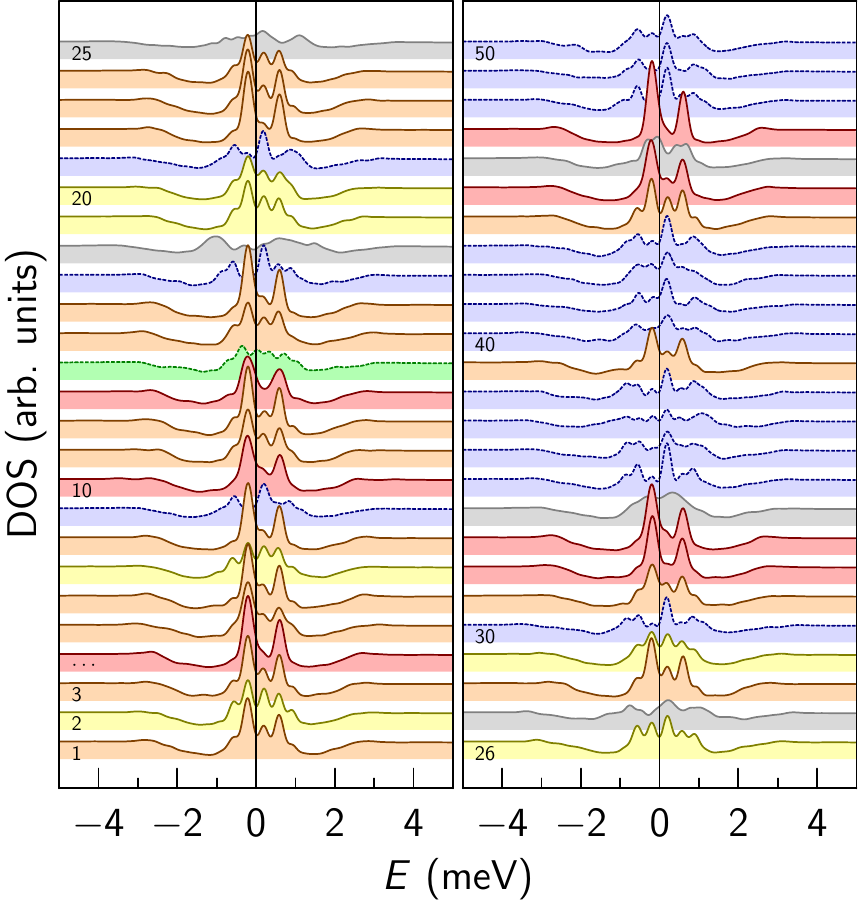}
\caption{\label{fig:fig8}
LDOS at the center (phase-singularity point) of 50 vortices sorted from bottom to top and from left to right by order of increasing distance from the center of the STM field of view (Fig.~1d of Ref.~\onlinecite{Chen-2018}; see Fig.~\ref{fig:figA5} in Appendix~\ref{app:vortices}). The curves are shifted vertically for clarity. The resolution is 0.25~meV ($N=18\,000$).
}
\end{figure}
Figure~\ref{fig:fig8} shows a collection of LDOS curves calculated at the core of the 50 vortices closest to the center of the field of view. A remarkable diversity of behaviors is seen. The red curves show vortices where the LDOS has one tall peak at negative energy and one weaker peak at positive energy, similar to what is found in the isolated vortex and in the ideal vortex lattices. A shoulder is often seen at a small positive energy between the two peaks, like in Fig.~\ref{fig:fig7}. In the curves highlighted in orange, the shoulder develops into a peak and two additional features appear on the high-energy tails of the main peaks. This further evolves into the yellow vortices, which show five peaks of roughly equal heights. The blue-dotted lines define another class, characterized by a prominent peak at positive energy. Inspection of Fig.~\ref{fig:figA5} in Appendix~\ref{app:vortices} reveals that these vortices are often (but not always) in the neighborhood of an area with lower vortex density. A Lorentz force therefore pushes these vortices into the vortex-deficient area, leading to a polarization of the bound states in that direction \cite{Berthod-2013a}. Since a prominent peak at positive energy is characteristic of the LDOS four to seven lattice parameters away from the vortex center (see Figs.~\ref{fig:fig3} and \ref{fig:figA4}), a possible interpretation of the blue curves is that, in these vortices, the singularity point where the LDOS was calculated is four to seven lattice parameters away from the ``electronic'' center of the vortex, where the low-lying bound states have their maximum. If that is the case, there must be another point inside the vortex where the LDOS resembles the red curves. Similarly, the orange and yellow curves are somewhat alike the LDOS two to three lattice parameters away from the center of an isolated vortex. Still, there are spectral shapes, like the yellow curves with peaks of almost equal heights and especially the green curve with a peak at zero energy, that are not seen close to an isolated vortex. Sorting out what stems from LDOS polarization---i.e., more or less rigid displacement of the LDOS relative to the phase-singularity point---from what would be genuinely new spectral shapes induced by positional disorder, requires a detailed analysis of the LDOS in the neighborhood of each vortex. This is left for a future study, and will hopefully also explain the gray curves, which seem to have very mixed up spectral shapes.

\section{Discussion}
\label{sec:discussion}

\begin{figure*}[tb]
\includegraphics[width=\textwidth]{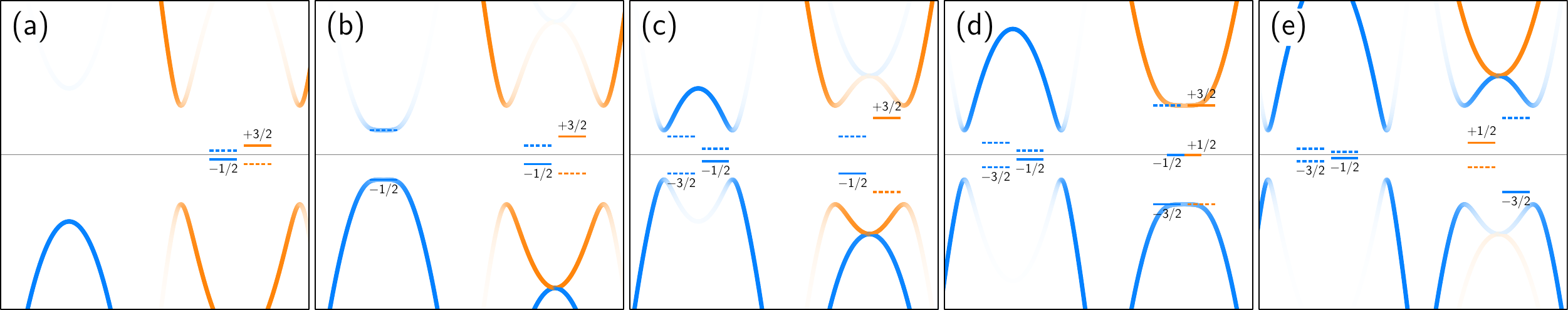}
\caption{\label{fig:fig9}
Evolution of the vortex bound states across two topological transitions. The bands at the $\Gamma$ and M points are shown schematically in the left and right sides of each panel and colored according to the BCS spectral weight. The pairs of horizontal bars indicate vortex levels; only the levels marked by a solid bar have weight at the vortex center. Numbers indicate the principal angular momentum. (a) Situation corresponding to the topmost curve in Fig.~\ref{fig:fig6}(a). (b) Appearance of the hole pocket at $\Gamma$. (c) Our model for \FeTeSe{}. (d) Transformation of the electron pocket at M to a hole pocket. (e) Situation corresponding to the topmost curves in Fig.~\ref{fig:fig6}(b).
}
\end{figure*}

Let us summarize the results first. A four-band tight-binding model (two inequivalent sites with two orbitals per site) with a bimodal uncorrelated random disorder and $s^{\pm}$ pairing symmetry explains well the STM topography recorded on cleaved \FeTeSe{} crystals. In order to explain the STM spectroscopy as well within this model, we need to assume gaps a factor of two smaller that those observed in ARPES and optical spectroscopy. While the disorder model captures well the spatial fluctuations of the STM spectra, it gives a spectral function much sharper than seen in ARPES experiments. The calculated topography shows some degree of correlation with the disorder, but the spectral gap (separation between coherence peaks) is uncorrelated with the disorder, even if the superconducting gap (order parameter) is allowed to adjust self-consistently in the disorder. The self-consistent order parameter of an isolated vortex, solved without disorder, is almost perfectly isotropic (cylindrical symmetry). The LDOS in the vortex shows resolution-limited peaks consistent with discrete levels. A spectral decomposition allows us to extract energies and wave functions from the calculated LDOS and to project the wave functions on angular-momentum eigenstates. We find that the bound states come in pairs at each angular momentum and can mix several angular momenta. The bound states change from being hole-like (negative angular momentum with the hole part of the wave function finite in the core) to being electron-like (positive angular momentum with the electron part of the wave function finite in the core) at a crossover energy. The hole-like states have higher energy in attractive Se-rich regions and lower energy in repulsive Te-rich regions, while electron-like states behave the opposite way. These variations are minute, though. Relatively large changes of the chemical potential are needed in order to substantially modify the vortex spectra, in ways that do not resemble the variations seen in the experiments. Finally, the LDOS at the phase-singularity point of a given vortex can present a variety of shapes, depending on the positions of disordered nearby vortices.

The electron-hole crossover in vortex bound states raises the question as to where the ``gender'' of these states comes from. It is tempting to bind each series of bound states in Fig.~\ref{fig:fig4} to one of the Fermi-surface pockets. This fails, because both series start with a hole-like state. Figure~\ref{fig:fig6} contains further insight. It shows two different transitions: (i) at $\Delta\mu=13$~meV, when the hole pocket at $\Gamma$ disappears, and (ii) at $\Delta\mu=-7.8$~meV, when the electron pocket at M transforms into a hole pocket. At the transition (i), one hole-like state leaves the core by moving into the continuum: this suggests that one series of hole-like core states is indeed bound to the hole band at $\Gamma$. At the transition (ii), the number of core states is conserved but their gender changes: one electron-like state leaves the core to the continuum while one hole-like state leaves the continuum to the core, and at the same time one hole-like state crosses zero energy and becomes electron-like. This makes sense if the band at M carries two core states, one being hole-like and one being electron-like, irrespective of whether the Fermi surface is electron- or hole-like. A scenario is proposed in Fig.~\ref{fig:fig9}. The hole band at $\Gamma$ behaves as usual: it carries one series of states with angular momenta $\mu=-1/2, -3/2, \ldots$, only the first of which has weight in the core. These levels scale with the small gap and progressively fill the core as the chemical potential enters the band. The band at M carries two series of states with opposite genders that scale with the large gap. When the band is electron-like [Figs.~\ref{fig:fig9}(a)--\ref{fig:fig9}(c)], the lowest hole-like state has momentum $-1/2$ and the lowest electron-like state has momentum $+3/2$. The latter hybridizes with $+1/2$ and thus shows up in the core. As the band gets emptied, the $+3/2$ state moves to the gap edge, while the $-1/2$ state has a nonmonotonic dependence, first moving toward the gap edge, then toward zero energy (see Fig.~\ref{fig:fig6}). When the chemical potential hits the quadratic touching point [Fig.~\ref{fig:fig9}(d)], the $-1/2$ state crosses zero energy and has mixed gender, while the state $+3/2$ reaches the gap edge and a state $-3/2$ appears at the opposite edge. Finally, when the band is hole-like [Fig.~\ref{fig:fig9}(e)], the lowest electron-like bound state has momentum $+1/2$ and the lowest hole-like state has momentum $-3/2$, hybridizes with $-1/2$, and becomes visible in the core [topmost curves in Fig.~\ref{fig:fig6}(b)]. This interpretation implies that, in the situation of Fig.~\ref{fig:fig9}(a), two bound states of opposite genders are present in the core even if $\Delta\mu$ is increased to a point that the hole band is out of the game. We have checked that this is actually true, and it takes another topological transition at $\Delta\mu=+99.5$~meV, where the M pockets touch and transform into $\Gamma$ hole pockets, to recover a unique gender in the cores, hole-like in this case. Likewise, a topological transition at $\Delta\mu=-55.4$~meV transforms the situation of Fig.~\ref{fig:fig9}(e) into an electron-like system with pockets at $(\pi/2,\pi/2)$.

The vortex-core spectrum of the model [Fig.~\ref{fig:fig2}(b)] with one tall peak at negative energy and one smaller peak at positive energy bears some resemblance with the observations made in part of the \FeTeSe{} vortices \cite{Chen-2018}, especially considering that nearby vortices, temperature, and a finite experimental resolution smoothen the theoretical spectrum. The energies do not match, though. The experiment reports peaks at $-0.6$, $+0.45$, $+1.2$, and $+1.9$~meV. The model has peaks at $-0.09$, $-0.2$, $+0.39$, $+0.53$, $+0.69$, and $+0.84$~meV. In qualitative agreement, both series show more electron-like than hole-like states. In qualitative disagreement, the measurement shows a hole-like state that is not the lowest, while the model has all hole-like states lying below the electron-like states. We would like to point out that the experiment faces a difficulty in locating the precise vortex center. The same difficulty arises in the theory when disordered nearby vortices polarize the LDOS (Sec.~\ref{sec:disorder}). For example, the series of spectra in Supplementary Fig.~5e of Ref.~\onlinecite{Chen-2018} shows a tall peak at $-0.88$~meV and a small peak at $+0.35$~meV in the putative vortex center, as well as a peak at $-0.28$~meV that develops far from the core. Such a configuration---with the lowest-lying state localized farther away from the core than higher-energy states---is quite unexpected and impossible for a clean isolated vortex. We speculate that the actual vortex center may be closer to where the $-0.28$~meV state is localized. The calculations show that the chemical disorder may shift the vortex levels to some extent; however these shifts are way too small to resolve the discrepancy. The seemingly smaller bound-state energies in the model compared with the experiment brings us back to the disagreement between the ARPES and STM spectral gaps. Scaling the gaps by a factor of two in order to match ARPES would scale the core levels by a factor of four, giving numbers somewhat more alike the experimental ones. For reconciling such a scaled model with the zero-field STM conductance, a third band is required with a gap of $\sim1$~meV. This gap is unresolved in photoemission, but needed to explain the sharp edge at 1~meV seen in tunneling. Additionally, one would need to invoke a nonlocal tunneling process that hides the larger gap near the M point. How the conductance delivered by this specific tunneling path relates to the LDOS inside vortices is an open question. In any case, the model as it stands confirms that the peaks observed by STM correspond to discrete vortex levels. This establishes a sharp contrast with the parent compound FeSe, which is claimed to host nodal \cite{Okada-2017, Hardy-2018} or at least very anisotropic \cite{Sprau-2017, Liu-2018, Rhodes-2018} gaps and shows a broad signature in vortices, with no sign of discrete states \cite{Song-2011}. The ability to observe and identify the ``conventional'' vortex states is an asset for demonstrating the unconventional nature of the Majorana zero mode in vortices where both types of quasiparticles coexist \cite{Wang-2018, Sun-2017}.

The intricate mixing of electron- and hole-like bound states in vortices is worth investigating further. Our results indicate that working at low field to approach as much as possible the isolated-vortex limit may be a good strategy in order to address the genuine vortex signature. It is expected that the vortices appear less distorted under the STM at low field than at 5~T and that the maximum of the zero-bias conductance does coincide with the electronic vortex center, while at higher fields they can be separated. Very much like the Chebyshev expansion, the STM lacks a direct access to the energies and wave functions of the bound states. A spectral decomposition of the tunneling conductance, analogous to the LDOS decomposition used in the present work---if it turns out to be feasible---would be of great value for distinguishing nearly degenerate states and determining their quantum numbers. The reverse behaviors of electron- and hole-like states relative to the chemical disorder may also be used as an identification tool, by studying the statistics of a given bound state in many vortices and correlating it with the topography.

On the theory side, many questions remain. Beside the ARPES/STM gap problem already discussed, a more realistic model of disorder may be needed: as the cartoon used here has almost no effect on the vortex spectroscopy, it does not provide a clue as to why the vortices are pinned at disordered positions. Another intriguing issue is the connection between the symmetries of the vortex and those of the Fermi surface. In the quantum regime, it is expected that the breaking of cylindrical symmetry is led by the Fermi surface rather than by the gap anisotropy \cite{Berthod-2016, Uranga-2016, Berthod-2017}. In this work, we have deliberately made the Fermi pockets maximally symmetric by a choice of the parameter $t_{2d}$ (Sec.~\ref{sec:model}). It would be interesting to monitor the evolution of the insets in Fig.~\ref{fig:fig2}(b) as the parameter $t_{2d}$ is varied and the electron pocket gains a fourfold harmonics oriented either along $(\pi,0)$ or along $(\pi,\pi)$, depending on $t_{2d}$. A further extension of our model should also consider the Zeeman splitting. In a 5~T field, this amounts to an energy scale of 0.58~meV that is not small compared with the gap size, unlike for the cuprates at similar fields. Going beyond the two-dimensional idealization, or at least modeling the three-dimensional effects and understanding how they change the spectra, is also a direction to explore. Finally, it is necessary to clarify whether there exists a link between the electron- or hole-like nature of the vortex bound states and the charge of elementary carriers as measured by the Hall coefficient, which was sometimes found to change sign as a function of temperature \cite{Tsukada-2010, Zhuang-2014}.

\section{Conclusion}
\label{sec:conclusion}

The direct observation of individual Caroli--de Gennes--Matricon bound states has been a rare event since their prediction in 1964. It appears that the low-density nearly compensated metal \FeTeSe{} has just the right set of band-structure and pairing parameters to enable this observation with high-resolution and low-temperature STM. Yet, the multiband nature of the compound with mixed electron-hole character together with nonlocal pairing, not to mention the intrinsic chemical disorder, challenges the well-established understanding for a clean single-band metal with local pairing. The model studied here is a first step in trying to understand how the vortex bound states behave in this more complex setup and deep in the quantum regime. The energy scales involved and the requirement that the interlevel spacing due to finite-size effects be much smaller than the superconducting gap immediately prompts unusually large system sizes that defeat ordinary methods based on straight diagonalization. The Chebyshev expansion allows one to get through this, with the advantage of being a real-space method well suited for disordered systems. Our calculations give a number of original insights without providing a completely satisfactory description of the observations made in \FeTeSe{}. We hope that this will motivate follow-up studies to refine and clarify the experimental data and develop the theory further.

\begin{acknowledgments}
This research was supported by the Swiss National Science Foundation under Division II. The calculations were performed at the University of Geneva with the clusters Mafalda and Baobab.
\end{acknowledgments}

\appendix

\section{Energy resolution of the Chebyshev expansion}
\label{app:resolution}

One of the main advantages of the Chebyshev expansion (\ref{eq:LDOS}) is that it provides for the LDOS a formula that is an analytical function of energy. On the contrary, calculations of the Green's function performed by direct inversion of the Hamiltonian matrix must be repeated independently at each energy of interest. The exact LDOS is a sum of Dirac delta functions centered at the energy eigenstates, a quantity that is discrete for any finite-size system, and which Eq.~(\ref{eq:LDOS}) approximates by a continuous function where each delta function is replaced by a function $\delta_N(E)$ of finite $N$-dependent width. The function $\delta_N(E)$ equals the DOS calculated for a one-orbital Hamiltonian $H=0$, that is, with a single eigenvalue at zero energy. Since $T_n(0)=\cos(n\pi/2)$, the resolution function is
	\begin{multline}
		\delta_N(E)=\frac{1}{\pi\mathfrak{a}}\mathrm{Re}\,\left\{\frac{1}{\sqrt{1-(E/\mathfrak{a})^2}}
		\right. \\ \left. \times
		\left[1+2\sum_{n=1}^Ne^{-in\arccos(E/\mathfrak{a})}\cos\left(\frac{n\pi}{2}\right)K_n\right]\right\}.
	\end{multline}
This function is similar to a Gaussian, however with tiny oscillations in the tails. In order to evaluate its width, we calculate
	\begin{align}
		\nonumber
		\Delta E&=2\sqrt{2\ln(2)\int_{-\infty}^{\infty}dE\,E^2\delta_N(E)}\\
		&=2\mathfrak{a}\sqrt{\ln(2)(1-K_2)}\approx\frac{2\sqrt{\ln 4}\pi\mathfrak{a}}{N},
	\end{align}
where the approximate result is valid at large $N$. For a Gaussian, this definition yields the full width at half maximum. The resolution functions calculated for $N=9000$ and $N=18\,000$ are plotted in Fig.~\ref{fig:figA1}(b) in Appendix~\ref{app:gap}. The corresponding energy resolutions are $\Delta E=0.5$~meV and 0.25~meV, respectively.

\section{Comparison of the DOS for full and halved ARPES gaps}
\label{app:gap}

\begin{figure}[b]
\includegraphics[width=0.7\columnwidth]{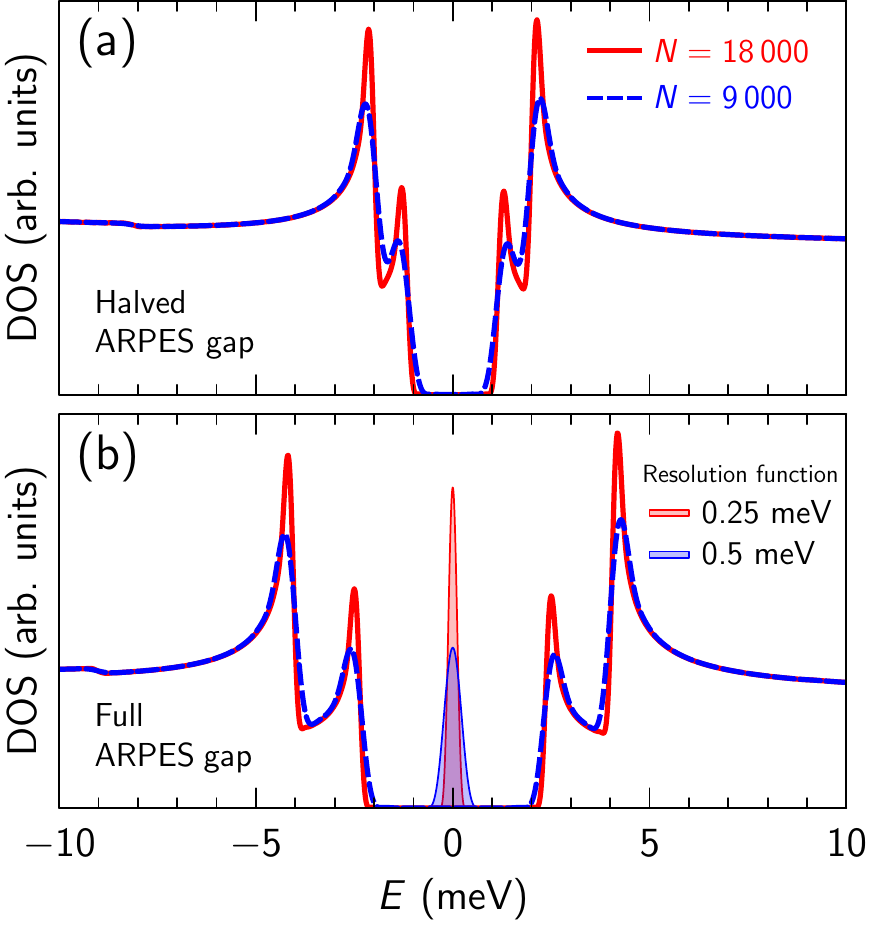}
\caption{\label{fig:figA1}
(a) DOS calculated without disorder ($V=0$) for an order parameter $\Delta_2=1.775$~meV and $\Delta_3=-0.475$~meV, that is, half the values measured by ARPES, and for two Chebyshev orders $N$. (b) Same calculation with the ARPES values $\Delta_2=3.55$~meV and $\Delta_3=-0.95$~meV \cite{Miao-2012}. The shaded curves show the resolution functions for each Chebyshev order.
}
\end{figure}

Figure~\ref{fig:figA1}(b) shows the DOS of the clean system (without the disorder associated with Te and Se potentials), calculated using the superconducting gap parameters deduced from ARPES measurements. Two gaps are clearly resolved, one of amplitude $\Delta_2\cos(k_{\mathrm{F},\Gamma}a)+\Delta_3\cos^2(k_{\mathrm{F},\Gamma}a)=2.4$~meV located on the $\Gamma$ pocket and one of amplitude $-\Delta_2\cos(k_{\mathrm{F},\Gamma}a)+\Delta_3\cos^2(k_{\mathrm{F},\Gamma}a)=-4.1$~meV located on the M pocket. The gap edges are square-root singularities as expected for an order parameter of $s$ symmetry, but appear rounded due to the finite resolution of the calculation truncated at order $N$. The amount of rounding can be compared with the resolution function, also displayed in Fig.~\ref{fig:figA1}(b). A structure near $-9$~meV marks the bottom of the electron band at $-7.8$~meV [see Fig.~\ref{fig:fig1}(b)], shifted by the gap opening. Figure~\ref{fig:figA1}(a) shows the DOS calculated with halved values of $\Delta_2$ and $\Delta_3$. The two gaps at 1.2 and 2~meV match the structures observed experimentally by STM \cite{Chen-2018, Wang-2018}. With a resolution of 0.5~meV ($N=9000$), the smaller gap still produces a small peak, while experimentally it appears more like a shoulder inside the main gap. For this reason, we have used $N=7000$ ($\Delta E=0.64$~meV) in our calculations of the LDOS in Fig.~\ref{fig:fig1}.

\section{Choice of the disorder strength}
\label{app:disorder}

\begin{figure}[b]
\includegraphics[width=0.7\columnwidth]{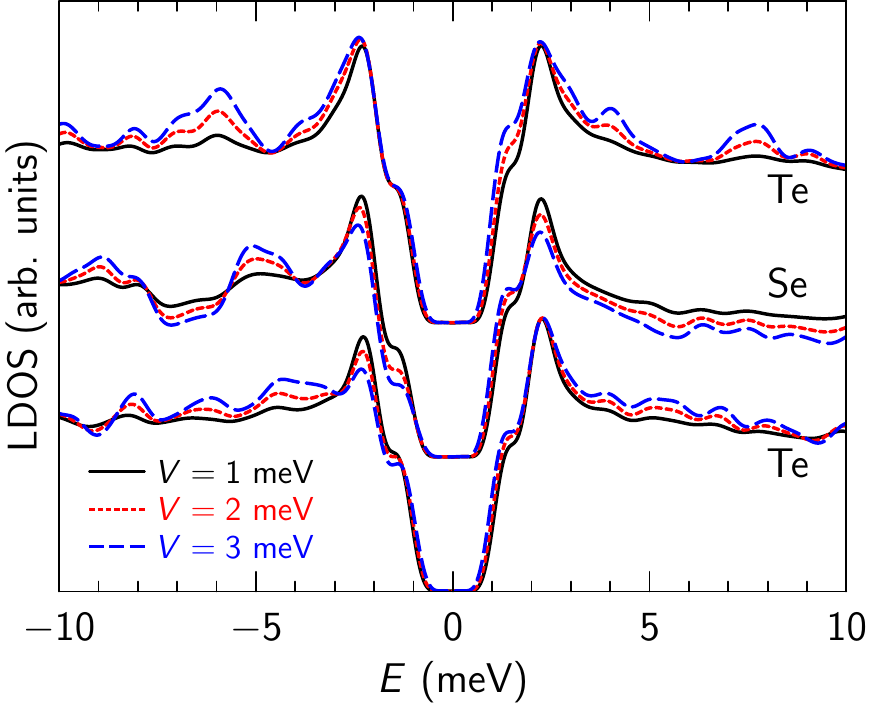}
\caption{\label{fig:figA2}
LDOS calculated at two sites with potential $V_{\mathrm{Te}}$ and one site with potential $V_{\mathrm{Se}}$ for three value of the potential strength $V$. The curves are shifted vertically for clarity. The energy resolution is 0.64~meV ($N=7000$).
}
\end{figure}

Figure~\ref{fig:figA2} shows how the LDOS depends on the disorder strength measured by the parameter $V$. We have picked three random sites on the lattice, two of which turning out to be Te sites and one a Se site. We recall that relative to the unperturbed chemical potential, the disorder potential is $+V$ at the Te sites and $-(11/9)V$ at the Se sites, in such a way that the spatially averaged potential is zero. One thus expects a shift of spectral weight towards positive energies at Te sites and towards negative energies at Se sites. This is clearly seen at $E>0$, where the LDOS is mostly increased by increasing disorder strength at the Te sites, and mostly decreased at the Se site. As the tunneling current is given by the integral of these LDOS curves from 0 to 10~meV, it is smaller at Se sites than at Te sites, which provides the contrast in Fig.~\ref{fig:fig1}(d). The disorder induces structures in the LDOS at energies that depend on the disorder configuration, but not on its strength. The amplitude of these structures scales with the strength of the disorder. We estimate that the typical amplitude of LDOS fluctuations for $V=1$~meV is similar to what is observed experimentally \cite{Chen-2018}. Figure~\ref{fig:figA2} also shows that the small gap on the $\Gamma$ pocket is more sensitive to disorder: there can be significant fluctuations of the coherence peak height at the edges of the small gap, while the fluctuations are weaker for the large gap. We note that for the calculations shown in Fig.~\ref{fig:figA2}, unlike for those shown in Figs.~\ref{fig:fig1}(d) and \ref{fig:fig1}(f), we neglected the self-consistent adjustment of the superconducting order parameter. Nevertheless, the tendency for the small gap to be more affected by disorder is confirmed by a statistical analysis of the peak height in the fully self-consistent map of Fig.~\ref{fig:fig1}(d). We find that (i) relative to the unperturbed DOS at the coherence-peak maximum, which marks the large gap, and at the energy of the shoulder, which marks the small gap, the average LDOS for $V=1$~meV ($V=2$~meV) goes down for the large gap by 1\% (5\%), while for the small gap it goes \emph{up} by 1\% (2\%); (ii) more importantly, the standard deviation of the LDOS distribution at these energies, relative to its average, is only 3\% (6\%) for the large gap for $V=1$~meV ($V=2$~meV), but as large as 8\% (15\%) for the small gap.

\section{Correlation of topography and gap map with disorder}
\label{app:correlation}

\begin{figure}[b]
\includegraphics[width=\columnwidth]{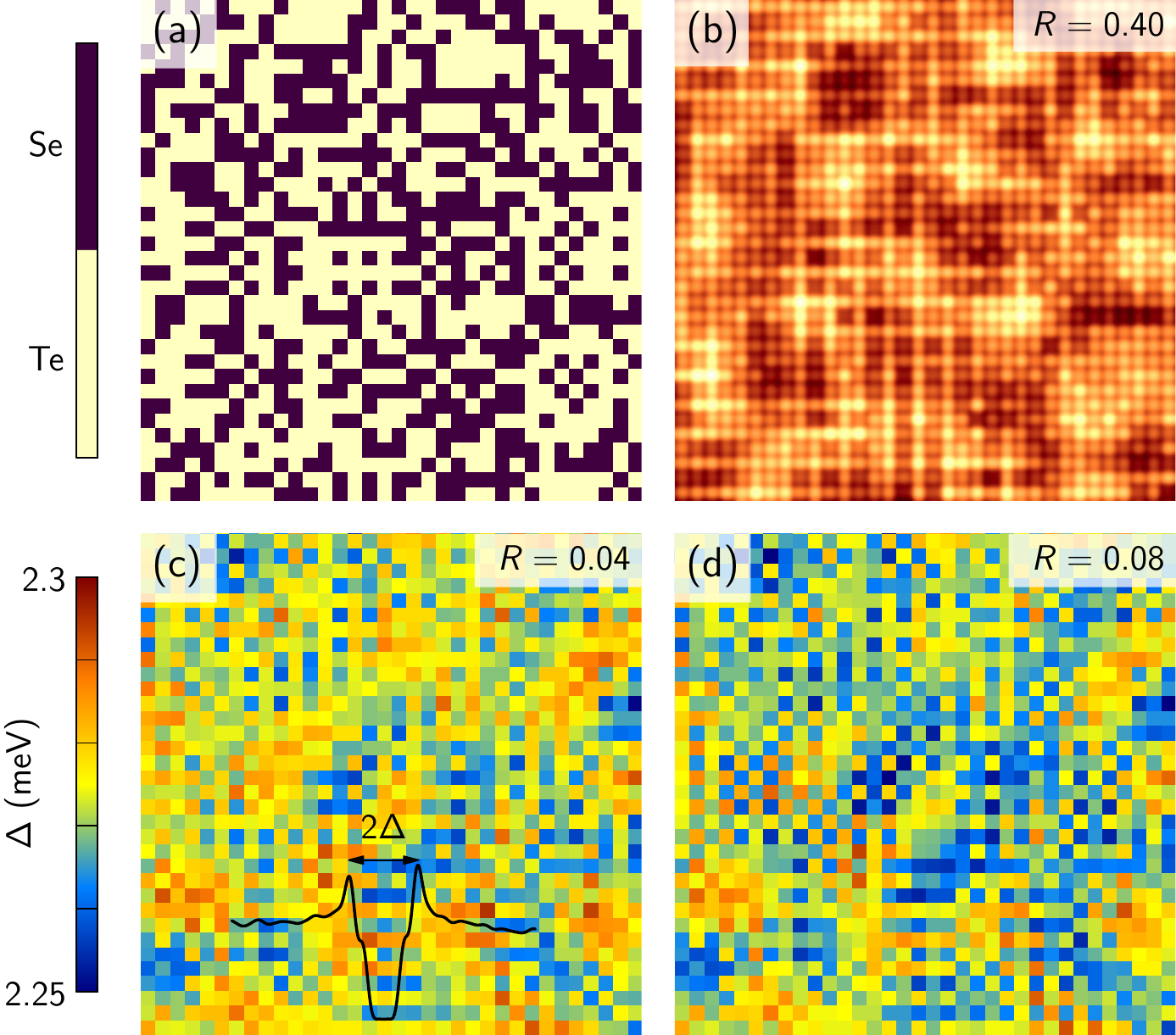}
\caption{\label{fig:figA3}
$35a\times35a$ spatial maps of (a) the potential landscape [same data as Fig.~\ref{fig:fig1}(c)], (b) the tunneling current [same data as Fig.~\ref{fig:fig1}(d)], (c) the non-self-consistent, and (d) the self-consistent spectral gap distributions. $R$ gives the correlation coefficient of each map with the disorder. The inset in (c) illustrates the definition of the spectral gap.
}
\end{figure}

In order to measure the correlation between the local tunneling current $I(\vec{r})$ shown in Fig.~\ref{fig:fig1}(d) and the disorder landscape shown in Fig.~\ref{fig:fig1}(c), we compute the correlation coefficient
	\begin{equation}
		R=\frac{\sum_{\vec{r}}[I(\vec{r})-\langle I\rangle][V_{\vec{r}}-\langle V\rangle]}
		{\sqrt{\sum_{\vec{r}}[I(\vec{r})-\langle I\rangle]^2\sum_{\vec{r}}[V_{\vec{r}}-\langle V\rangle]^2}},
	\end{equation}
where $V_{\vec{r}}$ is the local value of the potential, while $\langle I\rangle$ and $\langle V\rangle$ are the average current and potential, respectively. We find a value $R=0.4$, which quantifies the positive---although relatively weak---correlation between disorder and tunneling current that can be perceived by the eye. In Fig.~\ref{fig:figA3}, we compare this weak correlation with the absence of correlation between the potential and the spatial distribution of gap values. The spectral gap, defined as half the energy separation between the main coherence peaks [inset of Fig.~\ref{fig:figA3}(c)] is sensitive to disorder even if the order parameter is uniform. Figure~\ref{fig:figA3}(c) shows that the spectral gap, which takes the value 2.27~meV in the absence of disorder, varies between 2.25 and 2.3~meV when disorder is included without letting the order parameter adjust self-consistently. These variations show almost zero correlation with the disorder. With the self-consistent order parameter, the correlation is only marginally higher [Fig.~\ref{fig:figA3}(d)].

\section{ARPES structure factor}
\label{app:ARPES}

In a one-orbital system of noninteracting electrons, the spectral function is simply $A(\vec{k},E)=\delta(E-\xi_{\vec{k}})$, where $\xi_{\vec{k}}$ is the dispersion. Consequently, for any energy, the spectral weight is unity all along the constant-energy surface $\xi_{\vec{k}}=E$. The situation is different for multiorbital systems, where the spectral weight is a function of the wave vector $\vec{k}$. To obtain this function for our model, we start from the expression (\ref{eq:ARPES}) and note that, in the translation-invariant case, the Green's function can be expressed in terms of momentum eigenstates as
	\begin{equation}
		G_{\alpha'\alpha}(\vec{r}',\vec{r},E)=\langle\vec{r}'\alpha'|\sum_{\vec{k}\gamma=\pm}
		\frac{|\psi_{\vec{k}\gamma}\rangle\langle\psi_{\vec{k}\gamma}|}
		{E+i0-\xi_{\vec{k}}^{\gamma}}|\vec{r}\alpha\rangle.
	\end{equation}
In this expression, $\xi_{\vec{k}}^{\gamma}$ is the dispersion given by Eq.~(\ref{eq:xik}) and the momentum eigenfunctions are
	\begin{equation*}
		\langle\vec{r}\alpha|\psi_{\vec{k}\gamma}\rangle=\delta_{\alpha\gamma}\times\begin{cases}
		\psi_{\vec{k}\gamma}^{(1)}e^{i\vec{k}\cdot\vec{r}} & \vec{r}\in \text{first sublattice}\\[1em]
		\psi_{\vec{k}\gamma}^{(2)}e^{i\vec{k}\cdot(\vec{r}-\vec{\tau})} & \vec{r}\in \text{second sublattice}
		\end{cases}
	\end{equation*}
with $\vec{\tau}$ the vector joining the two sublattices---i.e., $\vec{\tau}=(a,0)$---and $\psi_{\vec{k}\pm}^{(1)}$, $\psi_{\vec{k}\pm}^{(2)}$ the values of the wave function on the first and second sublattice, respectively, in the unit cell sitting at the origin, for the eigenstate of energy $\xi_{\vec{k}}^{\pm}$. We insert these expressions into Eq.~(\ref{eq:ARPES}), where $M$ is to be understood as the total number of Fe sites, $M/2$ in each sublattice, and we split the expression into four terms, depending on whether $\vec{r}$ and $\vec{r}'$ belong to the first or second sublattice. If both $\vec{r}$ and $\vec{r}'$ belong to the first sublattice, we get $\frac{M}{2}\sum_{\alpha}\big|\psi_{\vec{k}\alpha}^{(1)}\big|^2\delta(E-\xi_{\vec{k}}^{\alpha})$. If both belong to the second sublattice, we get the same expression with $\psi_{\vec{k}\alpha}^{(2)}$ instead of $\psi_{\vec{k}\alpha}^{(1)}$. The normalization of the wave function is $\big|\psi_{\vec{k}\alpha}^{(1)}\big|^2+\big|\psi_{\vec{k}\alpha}^{(2)}\big|^2=\frac{2}{M}$, such that these two contributions together simply give $\sum_{\alpha}\delta(E-\xi_{\vec{k}}^{\alpha})$. Adding to this the contributions coming from $\vec{r}$ and $\vec{r}'$ being in different sublattices, we get
	\begin{equation}
		A(\vec{k},E)=\sum_{\alpha=\pm}\left[1+M\mathrm{Re}\,\psi_{\vec{k}\alpha}^{(1)}
		(\psi_{\vec{k}\alpha}^{(2)})^*e^{i\vec{k}\cdot\vec{\tau}}\right]\delta(E-\xi^{\alpha}_{\vec{k}}).
	\end{equation}
\begin{widetext}
The second term in the square brackets modulates the spectral weight along the constant-energy surfaces. After inserting the expressions of $\psi_{\vec{k}\pm}^{(1,2)}$, we are led to the final expression:
	\begin{equation}
		A(\vec{k},E)=\sum_{\alpha=\pm}\left\{1+\alpha\frac{t_1[\cos(k_xa)+\cos(k_ya)]}
		{\sqrt{t_1^2[\cos(k_xa)+\cos(k_ya)]^2+[2t_{2d}\sin(k_xa)\sin(k_ya)]^2}}
		\right\}\delta(E-\xi^{\alpha}_{\vec{k}}).
	\end{equation}
The function within curly braces vanishes for $k_y=0$ and $\alpha=-1$; hence the hole band has no spectral weight along $\Gamma$--M.
\end{widetext}

\section{Spectral decomposition of the vortex LDOS}
\label{app:decomposition}

\begin{figure}[tb]
\includegraphics[width=0.7\columnwidth]{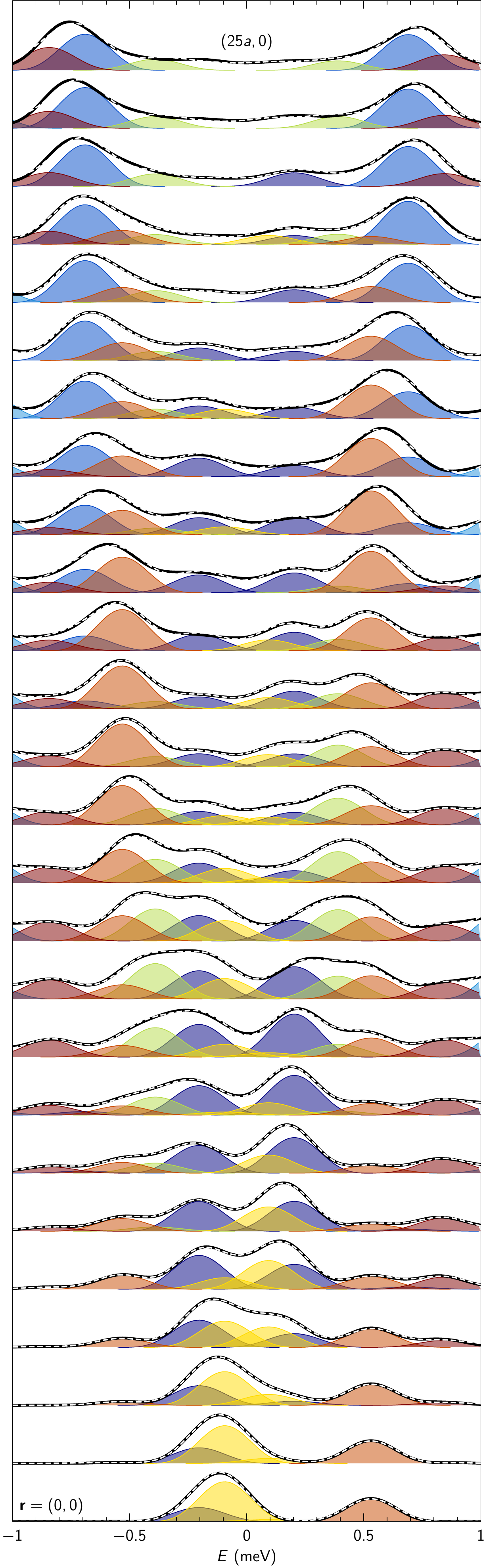}
\caption{\label{fig:figA4}
Series of LDOS spectra and their spectral decomposition.
}
\end{figure}

Figure~\ref{fig:figA4} displays a series of LDOS spectra (black lines) taken on a path going from the vortex center (bottom curve) to a distance of $25a$ along the (10) direction (top curve). All spectra are normalized to the peak height for clarity. The shaded curves show the spectral decomposition according to Eq.~(\ref{eq:LDOS2}); each shaded curve has the shape of the spectral-resolution function and pairs of peaks with the same color represent the electron and hole amplitudes of a given Bogoliubov excitation. The dashed lines show the sum of all peaks. Focusing one's attention on the black curves, one gets the impression that two peaks disperse in space as one moves away from the vortex center and end around $\pm0.8$~meV at $x=25a$. The spectral analysis shows how this apparent dispersion results from the superposition of non-dispersing peaks. This effect is well known in cases where the spectrum of bound states is dense (see, e.g., Ref.~\onlinecite{Gygi-1991}).

\section{Disordered configuration of vortices}
\label{app:vortices}

\begin{figure}[b]
\includegraphics[width=\columnwidth]{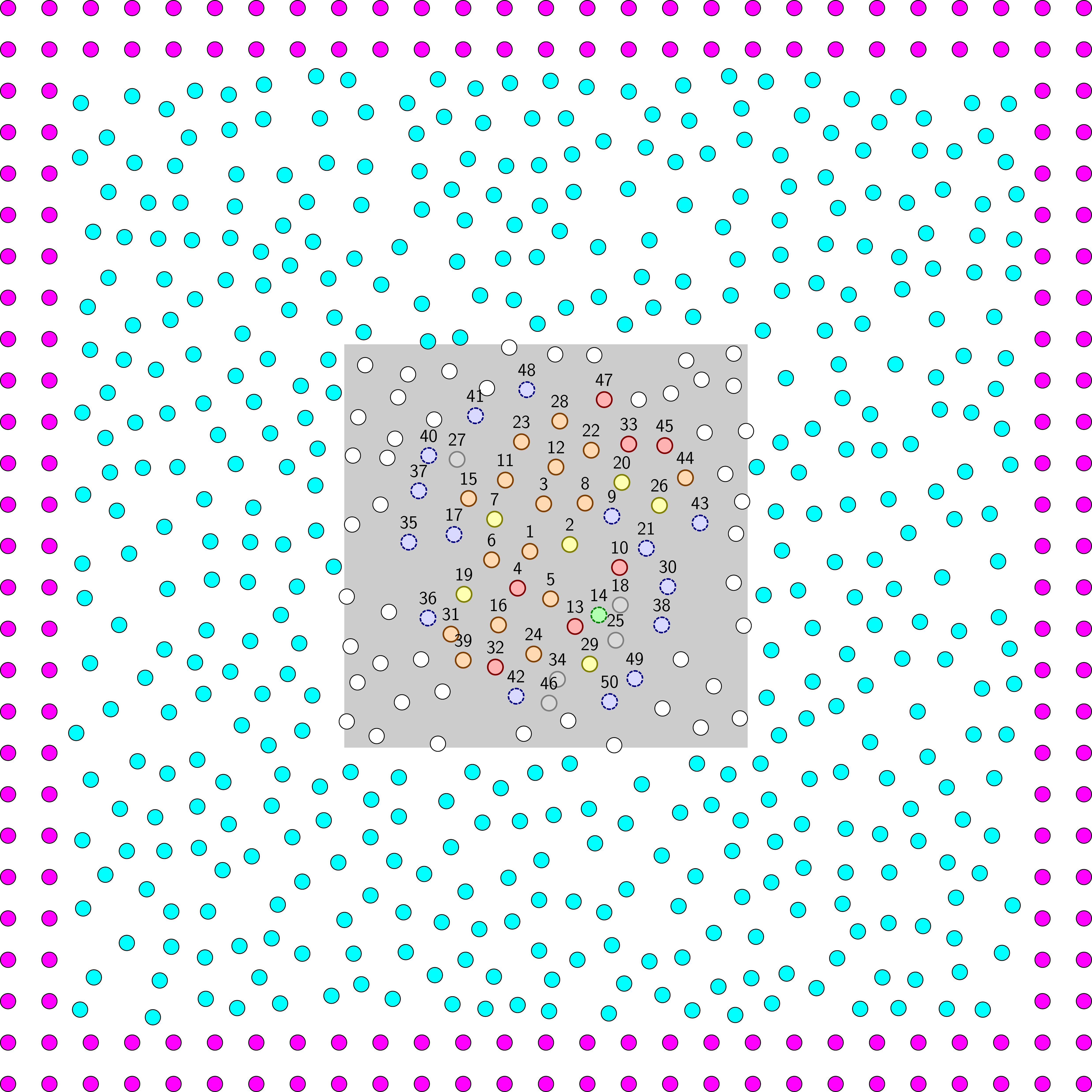}
\caption{\label{fig:figA5}
Vortex positions used for the simulations in Fig.~\ref{fig:fig8}. The numbers correspond to the spectra shown in Fig.~\ref{fig:fig8}.
}
\end{figure}

Figure~\ref{fig:figA5} shows the distribution of vortices used in the simulations. The gray square represents the 200~nm~$\times$~200~nm STM field of view of Fig.~1d in Ref.~\onlinecite{Chen-2018}. In this area, the vortex positions are located at maxima of the measured STM conductance (white and colors). There are 97 vortices, corresponding to a field of 5~T. The colored circles indicate the 50 vortices whose core LDOS is displayed in Fig.~\ref{fig:fig8} with the same color code. Outside the area, the vortex positions were generated randomly with the constraint of being at least 16~nm apart (cyan). The disordered vortices are surrounded by a regular square vortex lattice extending to infinity (magenta), which fixes the boundary condition for the phase of the order parameter \cite{Berthod-2016}.

\end{document}